\documentclass[fleqn,usenatbib]{mnras}

\usepackage[T1]{fontenc}
\usepackage{ae,aecompl}

\usepackage{newtxtext,newtxmath}

\usepackage[T1]{fontenc}

\DeclareRobustCommand{\VAN}[3]{#2}
\let\VANthebibliography\thebibliography
\def\thebibliography{\DeclareRobustCommand{\VAN}[3]{##3}\VANthebibliography}

\usepackage{graphicx}	
\usepackage{amsmath}	
\usepackage{amssymb}	

\usepackage{float}

\usepackage{chngcntr}
\counterwithout{table}{section}
\counterwithout{table}{subsection}
\counterwithout{table}{subsubsection}

\counterwithout{figure}{section}
\counterwithout{figure}{subsection}
\counterwithout{figure}{subsubsection}

\usepackage{wrapfig} 

\usepackage[]{hyperref}
\PassOptionsToPackage{hyphens}{url}
\usepackage{hyperref}

\usepackage{amsmath,amssymb}
\usepackage{wrapfig} 
\usepackage{chngcntr}
\counterwithout{figure}{section} 
\usepackage{graphicx}
\usepackage{float}
\usepackage{hyperref}
\usepackage{multicol}

\usepackage{ragged2e}
 
\usepackage{bbm}
\usepackage{dirtytalk} 
\usepackage[dvipsnames]{xcolor}

\title[Comparison of Sgr A* with GRMHD Simulations]{On the Comparison of AGN with GRMHD Simulations: I. Sgr A*}

\author[Richard Anantua, Sean Ressler and Eliot Quataert]{Richard Anantua$^{1,2,3}$\thanks{E-mail: ranantua@cfa.harvard.edu}, 
Sean Ressler$^{1,4,5}$, Eliot Quataert$^{1,4}$
\\
$^{1}$Astronomy Department, Theoretical Astrophysics Center, University of California, Berkeley, 601 Campbell Hall, \\$\ \ $Berkeley, CA 94720-3411, USA\\
$^{2}$
Center for Astrophysics $\vert$ Harvard \& Smithsonian, 60 Garden Street, Cambridge, MA 02138, USA
\\ 
$^{3}$ Black Hole Initiative at Harvard University, 20 Garden Street, Cambridge, MA 02138, USA
\\ 
$^{4}$ Department of Physics, 
University of California Berkeley, 1 LeConte Hall, Berkeley, CA 94720-3411, USA
\\
$^{5}$ Kavli Institute for Theoretical Physics, University of California Santa Barbara, Kohn Hall, Santa Barbara, CA 93107, USA
}

\date{Accepted XXX. Received YYY; in original form ZZZ}

\pubyear{2020}

\begin{document}
\label{firstpage}
\pagerange{\pageref{firstpage}--\pageref{lastpage}}
\maketitle

\begin{abstract}
We present models of Galactic Center emission in the vicinity of Sagittarius A* that use parametrizations of the electron temperature or energy density. These models include those inspired by two-temperature general relativistic magnetohydrodynamic (GRMHD) simulations as well as jet-motivated prescriptions generalizing equipartition of particle and magnetic energies. From these models, we calculate spectra and images and classify them according to their distinct observational features. Some models produce morphological and spectral features, e.g., 
image sizes, the sub-mm bump and low frequency spectral slope compatible with observations. Models with spectra consistent with observations produce the most compact images, with the most prominent, asymmetric photon rings. Limb brightened outflows are also visible in many models. 
Of all the models we consider, that which represents the current data the best is one in which electrons are relativistically hot 
when magnetic pressure is larger than the thermal pressure, but cold (i.e., negligibly contributing to the emission) otherwise.
 This work is part of a series also applying the \say{observing} simulations methodology to near-horizon regions of supermassive black holes in M87 and 3C 279. 
\end{abstract}




\section{Introduction}

Merely $d =$ 8.18 
kpc away \citep{2019arXiv190405721A} 
at the Galactic Center, Sagittarius A* (Sgr A*) is the best known accreting supermassive black hole. With $m_\mathrm{BH}=$ 4.14
million solar masses \citep{2019arXiv190405721A} 
(subtending 5.0 
$\mu$as at Earth), Sgr A* 
is a prime candidate for the next measurement of a black hole shadow by the Event Horizon Telescope (EHT,  \citealt{Doeleman2008})-- after the one in the giant elliptical galaxy M87. 
In April 2019, the EHT imaged emission around the central supermassive black hole of 
M87, finding a 42 $\mu$as-wide annulus with a Southern excess consistent with relativistic predictions of beamed emission in the Kerr metric of a 6.5 billion solar mass black hole  
\citep{2019ApJ...875L...1E}. The distribution of flux density over the EHT M87 image-- along with a conservative lower bound on the jet power of $L_\mathrm{M87}\ge 10^{42}$ erg/s-- precludes models with a non-spinning black hole, supporting the interpretation of the M87 jets as powered by the Blandford-Znajek mechanism \citep{BZ1977}. Image reconstruction of Sgr A* data has presented unique challenges relative to that for M87 due to the dynamical timescales of minutes as opposed to days \citep{2019ApJ...875L...1E}. GRAVITY \citep{Gillessen2010} has recently measured astrometrically the proper motion of several ``hot spots'' (i.e., near infrared flares) orbiting the Galactic Center black hole at $\sim 6$-$10$ gravitational radii ($M\equiv r_g = Gm_\mathrm{BH}/c^2$) \footnote{
$M$ for Sgr A* corresponds to $6.08\cdot 10^{9}$ m, 20.25 s and 4.96 $\mu\mathrm{as}$.}, suggesting that the innermost accretion flow may be relatively face on and strongly magnetized 
\citep{Gravity2018}. These horizon-scale probes of plasma physics, accretion physics, and strong general relativistic effects provide strong constraints on theoretical models.
 
Incorporating all these effects into analytic models is difficult, and thus general relativistic magnetohydrodynamic (GRMHD) simulations are the most promising avenues for theoretical modeling \citep{Gammie2003,Sadowski2014,White2016,Porth2017}.  These simulations solve the equations of MHD in the Kerr metric for a rotating black hole in 3+1 dimensions, naturally capturing the magnetorotational instability (MRI), the formation of jets, turbulence, and general relativistic effects.  

As an especially low-luminosity  active galactic nucleus (LLAGN)  with an observed Eddington ratio $\sim 10^{-9}$ \citep{Kataoka2018},
the dynamical time for accretion in Sgr A* is much shorter than the electron-ion Coulomb collision time.  This means that the flow can roughly be described as a two-temperature plasma \citep{Mahadevan1997} and that information beyond that provided by the standard GRMHD equations is needed to constrain the electron temperature, a key parameter in emission modeling.  Many approaches have been taken in the Literature for setting the electron temperature.  Simple post-processing prescriptions that ``paint on'' electron temperatures, such as those that set a constant electron-to-proton temperature ratio \citep{Mosci2009} everywhere or that set the electron temperature to be some function of plasma parameters \citep{Shcherbakov2012,Mosci2013,Chan2015}, are useful for rapidly and directly connecting observations to model parameters. More sophisticated treatments directly evolve the electron temperature alongside the single-fluid equations of GRMHD and incorporate knowledge of electron/ion heating gained from particle-in-cell (PIC) simulations of collisionless plasmas \citep{Ressler2015,Ressler2017,Sadowski2017,Chael2018}.  Though these models are more physically motivated and self-consistent, they are more expensive to run which makes it difficult to fully explore the parameter space of electron temperature spanned by our uncertainty in the dominant mechanism of electron/ion heating. Furthermore, conservative GRMHD simulations are unable to properly model the thermodynamics of flows in which the magnetic energy density far exceeds the rest mass energy density (as it often does in the jet).  Thus, two-temperature simulations that rely on an accurate calculation of the total fluid heating rate cannot be trusted in these regions; parametric models that depend only on the more reliable local magnetic energy density may be more appropriate there.\footnote{This may be part of the reason why two-temperature GRMHD simulations of Sgr A* have thus far under-produced the low frequency radio emission of Sgr A* \citep{Ressler2017,Chael2018}, as that emission is often assumed to be powered by the magnetized outflow. Alternatively, a small population of non-thermal particles can also explain the $\sim$ flat low frequency radio spectrum \citep{Ozel2000,Yuan2002}.} 

Given this uncertainty, we consider it instructive to further explore the dependence of the spectra and near-horizon-scale images of Sgr A* on different post-processing models of the electron temperature. In doing so, we include not only some of the prescriptions used in past work to model the Galactic Center, but also those attempting to mimic the qualitative behavior seen in recent two-temperature GRMHD simulations and those successfully used to model other systems, (e.g., \citealt{BK1979,Blandford2017,Anantua2018,Anantua2019}). We synthesize our results by categorizing the synthetic images and spectra into a small number of groups according to their emission physics.  This program adds to the foundation of efforts at classifying and understanding jet (or outflow)/accretion disk/black hole (JAB) emission through a unified framework flexible enough to model disk, corona and outflow or jet regions with a small set of parameters-- a methodology we call ``Observing" JAB Simulations. 

This work is a small addition to a vast literature on postprocessing simulations to understand JAB systems in the low accretion limit. 
For example, \citet{Dexter2009} simulated Sgr A* visibility amplitudes concordant with observed VLBI using $30^\circ$ inclination angle from the black hole spin axis and generated synthetic light curves with 30 min rise times and up to $50\%$ flux modulation.
\cite{Gold2017} compare synthetic Stokes maps of turbulent versus ordered magnetically arrested disk 
B-field configurations to EHT polarization data and favor the latter to account for the morphology and degree of linear polarization.
GRMHD simulations in \cite{Mosci2014} explain the Sgr A* radio spectrum and image size primarily by efficient electron heating in the outflow plasma.
\cite{Ryan2018} use general relativistic radiative magnetohydrodynamic (GRRMHD) simulations to model the M87 inner accretion flow and find inverse Compton cooling has a non-negligible backreaction within $10M$. The present work systematically considers simple parametric emission models to directly link microphysical processes to discrete observational signatures in Sgr A* and other AGN.

This paper is structured as follows: Section 2 is a brief synopsis of the observations of Sgr A* used to constrain our models; Section 3 describes the 
numerical simulations used in this work; Section 4 presents models for the accretion flow, electron temperature and emission; Section 5 provides results: images, spectra and light curves for our models; Section 6 compares the models in three different simulations to each other and to observations, resulting in a classification distilling various model morphologies, sizes and spectral shapes into four types; and Section 7 concludes.

In what follows, the speed of light, $c$, and the Boltzmann constant, $k_B$, are set to unity. Charge neutrality requires the electron number density, $n_e$, to be $\approx$ the proton number density, $n_p$ $\equiv n_0$. Then, the mass density is $\rho\approx \rho_p=m_p n_0$, where $m_p$ is the mass of a proton. Setting $m_p$ to 1 gives $P_e=\rho T_e$. The total temperature, $T_\mathrm{tot} = T_e + T_p $ is given by the simulation; $T_e$ is modeled.

\section{Observations}

Sgr A* has been observed for many years now in the radio, mm, infrared, and X-rays (
\citealp{Fish2011, Neilsen2013,Bower2015, Gravity2018}).  Time variability tends to increase with frequency, with the radio emission being the most stable and the infrared and X-ray emission showing frequent occurrence of large amplitude flares, likely caused by nonthermal particle acceleration.  Since we consider only thermal emission in this work, we treat the 10\% of the quiescent X-ray flux estimated to originate close to the black hole \citep{Neilsen2013} as an upper limit.  For a comprehensive list of the observational data points used in this work as constraints on our models (and plotted alongside our model spectra), see
\citet{Ressler2017}.

We use a 2D Gaussian semi-major axis size constraint of 49-63 $\mu$as (9.9-12.7$M$) incorporating long baseline data  \citep{Johnson2015,Lu2018} in an elliptical ring interpretation of the image to constrain our model image size. 
At 3.5 mm (86 GHz), \citet{Issaoun2019} ALMA observations give an intrinsic source axial sizes of $100\pm\ 18\mu$as (16.5-23.8$M$) and $120\pm\ 34\mu$as (17.3-31.0$M$). Recent Sgr A* measurements of non-zero closure phase, e.g., $5.0^{\circ +12.9^\circ}_{\ -4.6^\circ}$ from measurements along the SMT-CARMA-APEX triangle \citep{Lu2018}, rule out a spherically symmetric emission profile. 

\section{Principal GRMHD Simulations}

We perform a set of three simulations using the conservative, 3D, ideal GRMHD code {\tt HARM} \citep{Gammie2003}, which we denote as Standard and Normal Evolution (SANE), Magnetically Arrested Disk (MAD) and semi-MAD, all of which are endowed with electron temperature evolution based on \citet{Ressler2015}. The simulations all have dimensionless spin $a=0.5$, start from a \citet{FM1976} torus, and have resolutions of 320 x 256 x 64, uniform in the coordinates $x_1 (r,\theta), x_2 (r,\theta)$, and $x_3\equiv \varphi$, that 
are ``cylindrified'' and hyper-exponentiated \citep{SashaMAD} versions of modified Kerr-Schild (MKS) coordinates (\citealt{McKinney2004}; using $h = 0.3$), a process which is described in Appendix B of \citet{Ressler2017}. MKS coordinates focus resolution towards the midplane of the simulation, the ``cylindrification'' process increases the angular width of cells with $r\lesssim 10 M$, while the hyper-exponentiation extends the radial extent of the grid to thousands of $r_g$ by rapidly increasing the radial size of cells at $r > 400 M$. The adiabatic index of the total gas is $\gamma = 5/3$ and the adiabatic index of the electrons is $\gamma_e=4/3$. 
Where the three simulations differ is in both the size of the initial torus and the geometry of the initial magnetic field contained within this torus.  The SANE initial torus has an inner boundary of $r_\mathrm{in}=12 M$ and pressure maximum at $r_\mathrm{max}=24 M$, while the semi-MAD and MAD initial tori have inner radii of $r_\mathrm{in} = 15 M$ and pressure maxima at $r_\mathrm{max} = 34.5 M$.   While all simulations start with magnetic field lines that form single loops, the size and shape of the loops varies as determined by the magnetic vector potential, $A_\varphi$.  $A_\varphi$ scales as $A_\varphi\propto \rho$ in the SANE case and $A_\varphi\propto r^{4}\rho^2$ in the semi-MAD case, both of which we normalize such that $\max (P)/\max (P_B)=100$.  The MAD vector potential is more complicated and is computed as described in \citet{SashaMAD} while we normalize it such that $\min (P/P_B) = 100$. The resulting steady-state, time-averaged magnetic flux threading the horizon, $\Phi_{BH}$, for each of the runs are $\ll 1$ $M \sqrt{\dot{m}c}$ (SANE), $\approx 40$ $ M \sqrt{\dot{m}c}$ (semi-MAD), and $\approx 50$ $M \sqrt{\dot{m}c}$ (MAD).

The simulations implement the numerical density floor prescription $u_e\ge 0.01u_g$ from \cite{Ressler2015}. 
The simulation grid concentrates 3D spatial resolution in the disk, 
allowing for non-axisymmetric turbulence, kink instabilities, etc. in the MHD flow. The code has been parallelized using 
message passing interface (MPI).
The physical units of length and time are set by the mass of the black hole. 

Our fiducial simulation is the semi-MAD run.  We use a fiducial   
simulation time $T=10,000M$ to compare images and spectra.  Time-averaged $\dot{m}$ and $\dot{E}$ are found to be nearly constant in radius for the inner $r< 35M$ for SANE, semi-MAD and MAD simulations alike, indicating that inflow and outflow equilibrium is obtained for the regions of interest in this work. In fact, the MAD simulation is in equilibrium up to $r\lesssim 100M$.

\section{Emission Models}

\subsection{Electron Thermodynamics Models}
\label{sec:models} 

\subsubsection{Electron Evolution Model with Turbulent Heating}

All of our simulations also include an electron entropy equation as described in \citet{Ressler2015} (neglecting electron conduction), using the \citet{Howes2010} heating prescription for turbulent heating in collisionless plasmas.
We refer to this as the \say{Electron Evolution Model with Turbulent Heating} (or \say{Electron Evolution Model} for short). Note that the \citet{Howes2010} heating prescription is the result of calculations of the Landau damping of turbulence -- different dissipation mechanisms at the end of a turbulent cascade could produce different results.

The \citet{Howes2010} heating function is strongly dependent on plasma $\beta=P_g/P_B$, with a sharp transition between electrons being preferentially heated at $\beta \lesssim 1$ to protons being preferentially heated at $\beta > 1$. A direct consequence is that the relativistically hot electrons are confined to the coronal and jet regions of the simulations while the electrons in the midplane of the disk are non-relativistically cold.  

\subsubsection{Critical Beta Electron Temperature Model 
}

In an attempt to mimic the behavior of the Electron Evolution Model-- without explicitly including  
turbulent heating in an 
entropy equation-- we construct a post-processing function for the electron-to-total temperature ratio:
\begin{equation}
\frac{T_e}{T_\mathrm{tot} 
} = f e^{-\frac{\beta}{\beta_c}},
	\label{eq:ETempModel}
\end{equation}
where $0<f<1$ is a constant, and $\beta_c$ is the critical value of $\beta$ that approximately sets a maximum $\beta$ contributing to emission. 
We call this model the \say{Critical Beta Electron Temperature Model} (or \say{Electron Temperature Model} for short). 
 
This model converges to unique values for the ratio $T_e/T_p$ in the limits $\beta\to 0$ and $\beta\to \infty$ 
similarly to models in \citet{Mosci2014,Davelaar2018} employed by the EHT for M87 in which the ratio of proton-to-electron temperature is bounded by constants $R_\mathrm{low}$ and $R_\mathrm{high}$. 

\subsubsection{Constant Electron Beta Model}
 
Another viable post-processing prescription for the electron temperature is one in which the electron energy density $u_e$ is some fixed fraction of the magnetic energy density $u_B$.  This may be reasonable if magnetic reconnection is the dominant source of electron heating as it is presumed to be in jet regions.  This \say{Constant Electron Beta Model} (or \say{Constant $\beta_e$ Model}) is described by a single parameter through the relation $\beta_e=P_e/P_B=(\gamma_e-1)u_e/(b^2/2)=\beta_{e0}$ (constant), or:
\begin{equation}
P_e = \beta_{e0}P_B.
	\label{eq:BetaModel}
\end{equation} 
where $b^2\equiv b^\mu b_\mu$.

Note that equipartition of particle and electromagnetic energies corresponds to $\beta_{e0} \sim 1$. 
Models coupling near-equipartition jets to Sgr A*'s accretion flow have previously been examined in  \cite{Falcke1993} and \cite{Falcke2000}. There, the jets have been invoked to explain the radio emission and, 
in some models, the higher frequency spectrum. 

\subsubsection{Magnetic Bias Model}

We can generalize the Constant $\beta_e$ Model so that the electron pressure scales as powers of the magnetic pressure
\begin{equation}\label{BiasModel}
P_e=K_nP_B^n \sim b^{2n}
\end{equation}
where
\begin{equation}\label{eq:BiasModel}
K_n=K_1\frac{\langle P_B \rangle}{\langle P_B^n\rangle}=2^{n-1}K_1\frac{\langle b^2 \rangle}{\langle b^{2n}\rangle } 
\end{equation}
and $\langle \rangle$ denotes an average over cylindrical radii $2M<R<20M$ as in Appendix \ref{TablesAppendix} Table \ref{OldAvgbToNCylinder}. We call this the \say{Magnetic Bias Model} (or \say{Bias Model} for short).

Note that $n = 1,K_1\equiv\beta_{e0}$ corresponds to the Constant $\beta_e$ Model. By default, for $n\neq 1$, we take $K_1=1$ in this work in the interest of space, although there is a priori no strong motivation for a particular value of $K_n$.

\subsection{Radiative Transport}

We compute (mainly 230 GHz) images using the ray-tracing scheme {\tt IBOTHROS} \citep{Noble2007}, which includes the effects of synchrotron emission and absorption, while we compute spectra using the Monte-Carlo-based {\tt GRMONTY} \citep{Dolence2009}, which includes the effects of synchroton emission, absorption, and inverse Compton scattering.  For the purposes of radiative transport, we exclude regions of the simulation with $\sigma \equiv b^2/\rho>1$ where the GRMHD solution becomes less reliable. The pixel resolution used to produce IBOTHROS images in this work is 1$M$ per pixel in both directions. 

Since GRMHD simulations are scale free, for each electron temperature model we choose the physical mass unit such that the flux at 230 GHz matches the 2.4 Jy measurement \citep{Doeleman2008} (to $<2\%$).  
Table \ref{MUnit230GHz} lists the resulting mass accretion rate for each of our models. Note that negative accretion rate corresponds to inflow. 

Our fiducial viewing angle will be 90$^\circ$ (edge-on disk) and our fiducial observer frequency will be $\nu_\mathrm{Obs}=230$ GHz, 
 though we also consider 0$^\circ$ (face on disk), 45$^\circ$ and $\nu_\mathrm{Obs}=140$ THz, the latter for comparison with near infrared observations. In the absence of other labels, all of the images and  spectra shown correspond to fiducial values of viewing angle and frequency and to the semi-MAD simulation.

\begin{table*} 
	\centering
	\caption{Mass accretion rate yielding $\approx$ 2.4 Jy 230 GHz flux for our models. Negative values of mass accretion rate correspond to infall. The monochromatic flux is summed over $50M$ x $50M$ image regions. For comparison, the Eddington accretion rate for Sgr A*'s mass is $-0.09 \, M_\odot/\mathrm{yr}$.
	}
	\label{MUnit230GHz}
	\begin{tabular}{lcccr} 
		\hline
		Model & 
        &  Accretion Rate $M_\odot/\mathrm{yr}$ &
        \\
     & SANE
     & semi-MAD
     & MAD
     \\
     		\hline
		Electron Evolution with Turbulent Heating  
          & $-2.61\cdot 10^{-8}$ 
          & $-1.70
          \cdot 10^{-8}$ 
          & $-1.44\cdot 10^{-11}$
          \\
		\hline
		 Critical Beta Electron Temperature ($f,\beta_c$) &  
        & \\
        ($0.1,0.01$)& $-1.11\cdot 10^{-7}$ 
       
        &  
        $-1.72
        \cdot 10^{-7}$
        & $-1.90\cdot 10^{-9}$
        \\
       ($0.1,0.1$)& $-8.10\cdot 10^{-8}$ 
       & 
       $-1.16
       \cdot 10^{-7}$
       & $-4.31\cdot 10^{-10}$
       \\
       ($0.1,1.0$)& 
       $-6.92
       \cdot 10^{-8}$ 
       & 
       $-9.14
       \cdot 10^{-8}$
        & $-2.58
       \cdot 10^{-10}$
       \\
       ($0.5,0.01$)& $-2.57\cdot 10^{-8}$
      
       &  
       $-3.62
       \cdot 10^{-8}$
       & $-3.11\cdot 10^{-10}$ 
       \\
       ($0.5,0.1$)& $-1.52\cdot 10^{-8}$
       &  
       $-1.13
       \cdot 10^{-8}$
       & $-7.87\cdot 10^{-11}$
       \\
       ($0.5,1.0$) & $-3.93
       \cdot 10^{-9}$
       & 
       $-6.06
       \cdot 10^{-9}$
       & $-2.15
       \cdot 10^{-11}$
       \\
        \hline
		Constant Electron Beta ($\beta_{e0}$)&  &  
		\\
         0.01   & $-4.75\cdot 10^{-9}$
         & 
         $-2.92
         \cdot 10^{-9}$ 
         & $-3.14\cdot 10^{-11}$  
         \\
         0.1   & 
          $-3.11\cdot 10^{-10}$
           &          
         $-6.28
         \cdot 10^{-10}$ & $-4.83\cdot 10^{-12}$
         \\
         1.0   & 
         $-3.90\cdot 10^{-11}$ 
          &
         $-3.86
         \cdot 10^{-10}$
         & $-1.62\cdot 10^{-12}$
         \\
        \hline
		 Magnetic Bias ($n$)
         &  &  
         \\
		0 & $-4.04\cdot 10^{-9}$
        & 
        $-1.41
        \cdot 10^{-9}$ 
        & $-1.66\cdot 10^{-9}$
        \\
        1 & $-3.90\cdot 10^{-11}$
        & 
        $-3.86
        \cdot 10^{-10}$ 
        & $-1.62\cdot 10^{-12}$
        \\
        2 & $-2.21\cdot 10^{-11} $ 
        &  
        $-4.54
        \cdot 10^{-10}$ 
        & $-1.95\cdot 10^{-12}$ 
        \\
        \hline
	\end{tabular}
\end{table*}

\section{Results}

\subsection{Electron Temperature Profiles}
\begin{figure}
\begin{align}\nonumber
& \hspace{1cm}\includegraphics[height=170pt,width=190pt,trim = 6mm 1mm 0mm 1mm]{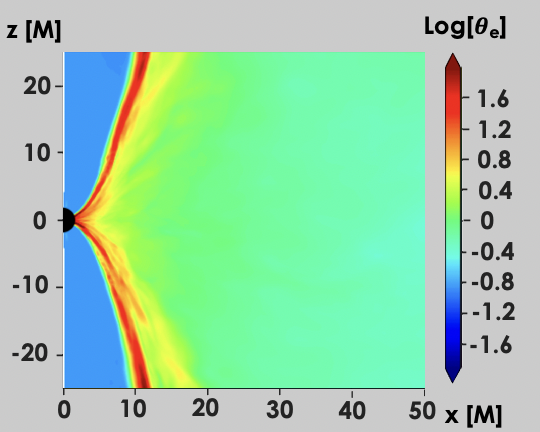} 
\end{align} 
\begin{align}\nonumber
&  \hspace{1cm}\includegraphics[height=170pt,width=190pt,trim = 6mm 1mm 0mm 1mm]{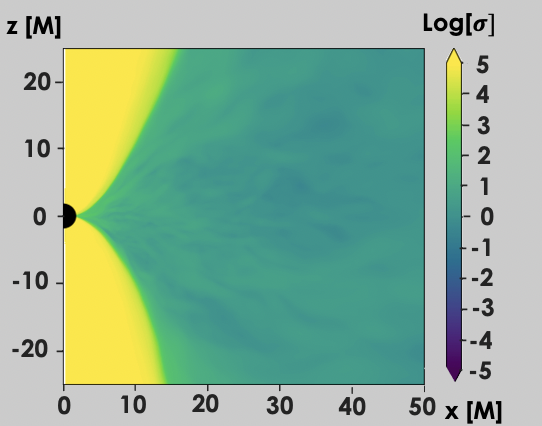}
\end{align} 
\caption{Slices of electron temperature $r\theta$ profile for the Electron Evolution Model (Top) and slice of $\sigma=b^2/\rho$ (Bottom). We use common log scales to present large dynamical ranges.
}\label{VerticalTeSlicesElectronEvolModelAndSigmaSlice}
\end{figure} 

\begin{figure}
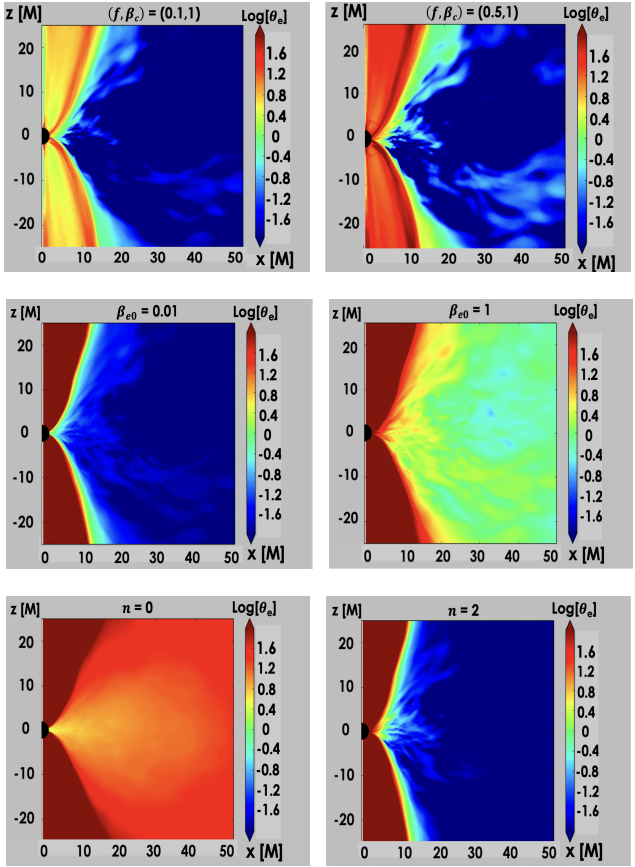

\begin{align}\nonumber
& \includegraphics[height=100pt,width=110pt,trim = 6mm 1mm 0mm 1mm]{LogElectronTempProfilefEqPt1BetaCEq1
.png} & \includegraphics[height=100pt,width=110pt,trim = 6mm 1mm 0mm 1mm]{LogElectronTempProfilefEqPt5BetaCEq1
.png}
\end{align}
\begin{align}\nonumber
  & \includegraphics[height=100pt,width=111pt,trim = 6mm 1mm 0mm 1mm]{TeProfileBetae0EqPt01
  .png} & \includegraphics[height=100pt,width=110pt,trim = 6mm 1mm 0mm 1mm]{TeProfileBetae0Eq1
  .png} &   
\end{align}
\begin{align}\nonumber
  & \includegraphics[height=100pt,width=110pt,trim = 6mm 1mm 0mm 1mm]{LogElectronTempProfilenEq0_
  .png} & \includegraphics[height=100pt,width=110pt,trim = 6mm 1mm 0mm 1mm]{LogElectronTempProfilenEq2
  .png} & 
\end{align} 
\caption{Electron temperature $r\theta$ profiles for the Critical Beta Electron Temperature Model with $(f,\beta_c)=(0.1,1)$ (Top Left) 
and $(0.5,1)$ (Top Right). Also, electron temperature $r\theta$ profiles for 
Constant $\beta_e$ Models with $\beta_{e0}=0.01$ (Middle Left), $\beta_{e0}=0.1$ (Middle Right), 
and Bias Models with $n=0
$  (Bottom Left) and $n=2
$ (Bottom Right). The temperature is expressed in dimensionless form $\theta_e=k_BT_e/m_ec^2$.
}\label{VerticalTeSlicesElectronTempAndBetaAndBiasModels}
\end{figure}

Figure \ref{VerticalTeSlicesElectronEvolModelAndSigmaSlice} shows the azimuthally averaged electron temperature distribution for the Electron Evolution Model and plasma electromagnetic-to-particle flux density $\sigma$ in the $r$-$\theta$ plane. The electron temperature is greatest on the boundary between the high $\sigma$ outflow and lower $\sigma$ outflow and inflow as found in \citet{Ressler2015}. Figure \ref{VerticalTeSlicesElectronTempAndBetaAndBiasModels} shows the azimuthally averaged electron temperature distributions in the $r$-$\theta$ plane resulting in each of our models for a few select parameter choices.
For Critical Beta Electron Temperature Models, a boundary layer on the disk-jet interface (mild outflow+corona) is the region with highest $T_e$. This behavior is consistent throughout the parameter space $f\in\{0.1,0.5\}$ and $\beta_c\in\{0.01,0.1,1\}$. 

The electron temperature profiles in various Constant $\beta_e$ Models in the middle panels of Figure \ref{VerticalTeSlicesElectronTempAndBetaAndBiasModels} are hottest for the strong interior outflow, or \say{spine,} characterized by low density and high magnetization. 
In Bias Models at the bottom panels of Figure \ref{VerticalTeSlicesElectronTempAndBetaAndBiasModels}, electron temperature is also highest near the coherent, electromagnetically-dominated outflow. Now, however, the  radial profile is strongly dependent on the exponent $n$, which enhances the variation of emission (a function of $b$) with cylindrical radius in the simulation. 
 

\subsection{Electron Evolution Model with Turbulent Heating} 

\subsubsection{Electron Evolution Model Images}

\begin{figure} 
\begin{align}\nonumber
  & \hspace{1.0cm} \includegraphics[height=150pt,trim = 6mm 1mm 0mm 1mm]{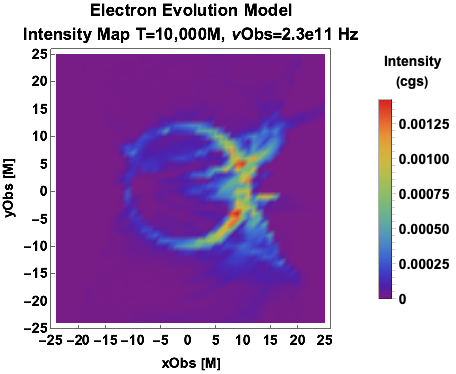}
  & 
\end{align}
\caption{Electron Evolution Model image at 230 GHz.}\label{ElectronEvolutionModel}
\end{figure}

The Electron Evolution Model image\footnote{
Each individual pixel in the image plane (normal to the observer) has an $I_\nu = d F_\nu / d\Omega$ in cgs units, and $F_\mathrm{tot}\approx2.4\cdot10^{-23}$erg s$^{-1}$ cm$^{-2}$ Hz$^{-1}$  = 2.4 Jy is just the sum of all the $dF_\nu$'s. Thus, multiply each cgs intensity colored pixel value by 
57.9 to get its flux density in Jy.} in Figure \ref{ElectronEvolutionModel} shows 
a ring of lensed emission around the black hole event horizon and a small outflow: 
the 
whispy 
outflow is visible at radii 
not exceeding $30M$
and is limb brightened. Asymmetry in the photon ring is apparent, due to Doppler shifts at the edge on viewing angle. The images in this paper are rendered 
in observer coordinates left-right and up-down inverted relative to \citet{Ressler2017}. 
In our implementation of a new imaging pipeline 
90$^\circ$ inclination has the accretion flow approaching on the right and black hole spin pointing down.

 \subsubsection{Electron Evolution Model Spectra}

\begin{figure}\nonumber
\begin{align}
\hspace{0.5cm}\includegraphics[height=180pt,width=210pt,trim = 6mm 1mm 0mm 1mm]{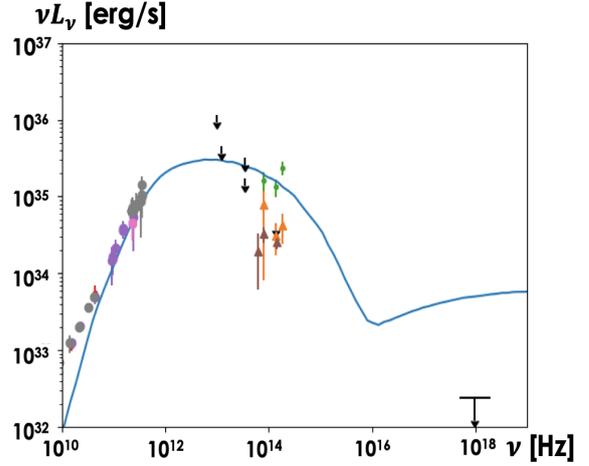}
\end{align} \caption{
Spectrum generated from the Electron Evolution Model with Turbulent Heating. 
}\label{ElectronEvolSpec}
\end{figure}

The Electron Evolution Model spectrum is shown in Fig. \ref{ElectronEvolSpec}. 
The infrared bump is well fit, however the model's low frequency slope steepens under the data around 10 GHz, and the model significantly overpredicts the X-ray emission. Note that this spectrum also appears in \citet{Ressler2017}.

 \subsection{Critical Beta Electron Temperature Model}

\subsubsection{Electron Temperature Model Images}

\begin{figure}
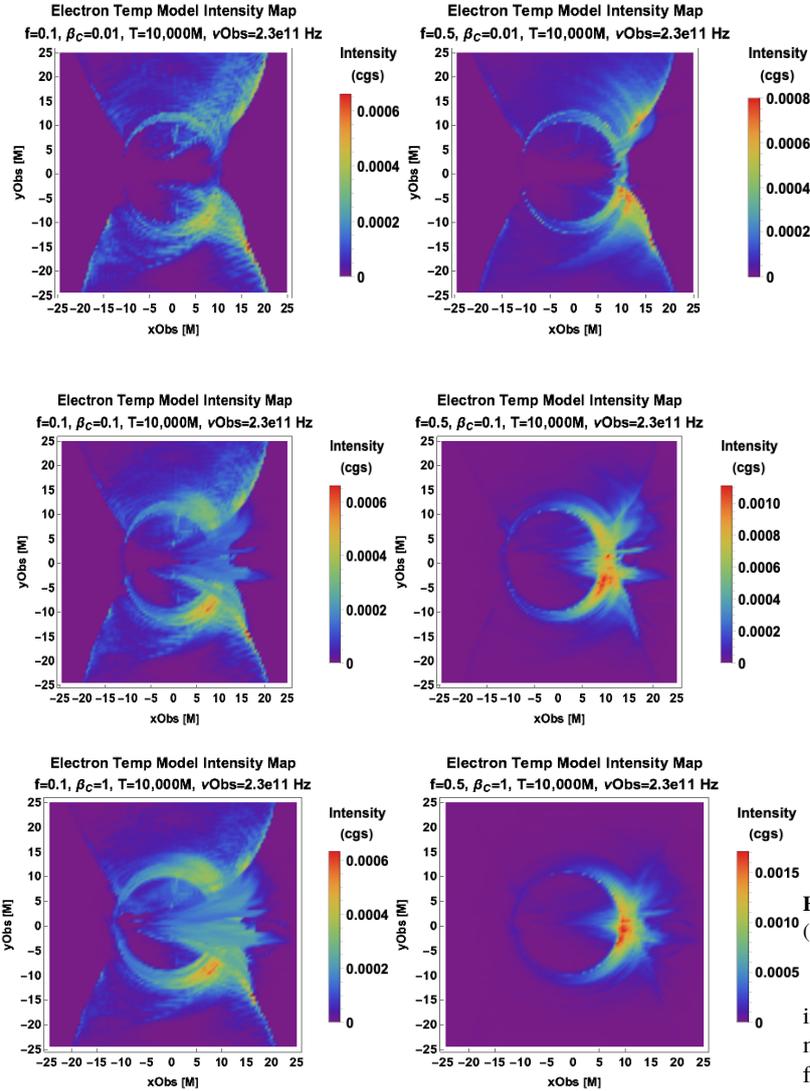
\nonumber
\begin{align}
\hspace{0mm} & \includegraphics[height=125pt,width=115pt,trim = 35mm 1mm 0mm 1mm]{PlotfEqPt1BetaTotCritEqPt01HighResZoomMunit1Pt11648e21Flux1Pt85762DX25DY25
.png} & \includegraphics[height=125pt,width=115pt,trim = 10mm 1mm 30mm 1mm]{PlotfEqPt5BetaTotCritEqPt01ZoomHighResMunit2Pt35543e20Flux1Pt99745DX25DY25
.png} & 
\end{align}
\begin{align}\nonumber
  \hspace{0mm} & \includegraphics[height=125pt,width=110pt,trim = 35mm 1mm 0mm 1mm]{PlotfEqPt1BetaTotCritEqPt1HighResZoomMunit7Pt58373e20Flux2Pt0889DX25DY25
  .png} & \includegraphics[height=125pt,width=110pt,trim = 10mm 1mm 30mm 1mm]{PlotfEqPt5BetaTotCritEqPt1ZoomHighResMunit7Pt3322e19Flux2Pt32099DX25DY25
  .png} & 
\end{align}
\begin{align} 
  \hspace{-1.5mm} &  \includegraphics[height=125pt,width=115pt,trim = 35mm 1mm 0mm 1mm]{PlotfEqPt1BetaTotCritEq1HighResZoomMunit5Pt94859e20Flux2Pt20837DX25DY25
  .png} & \includegraphics[height=125pt,width=115pt,trim = 10mm 1mm 30mm  1mm]{PlotfEqPt5BetaTotCritEq1HighResZoomMunit3Pt94188e19Flux2Pt34777DX25DY25
  .png} &  
\end{align} 
\caption{Critical Beta Electron Temperature Model images for $f=0.1$ (Left Panels) and $f=0.5$ (Right Panels) for $\beta_c=0.01$ (Top Panels), $\beta_c=0.1$ (Middle Panels) and $\beta_c=1$ (Bottom Panels).
}\label{ElectronTemperatureModelImages}
\end{figure}

The images in Fig. \ref{ElectronTemperatureModelImages} 
show the $\beta_c$ variation of the Electron Temperature Model for $f=0.1$ and $f=0.5$, respectively. 
For most of the parameter space, the Electron Temperature Model appears to be a fairly uniform projection of the inflow-outflow boundary/coronal region. For the highest values of $f$ and $\beta_c$, the images become more asymmetric and the outflow is limited to smaller radii.
Optical depth effects 
are apparent as follows: In the optically thin case, the prescription $T_e=fT_pe^{-\beta/\beta_c}$ at constant $\beta_c$ has line-of-sight intensity proportional to $f$; yet the brightness of intensity maps in Fig. \ref{ElectronTemperatureModelImages} are not simple re-scalings of each other as a function of $f$. 
This is most noticeable at the highest value $\beta_c=1$, where the intensity map appears several times more compact and lopsided as we vary $f$ from 0.1 to 0.5.

\subsubsection{Electron Temperature Model Spectra}

\begin{figure}\nonumber
\begin{align}
\hspace{1.5cm}\includegraphics[height=150pt,width=180pt,trim = 6mm 1mm 0mm 1mm]{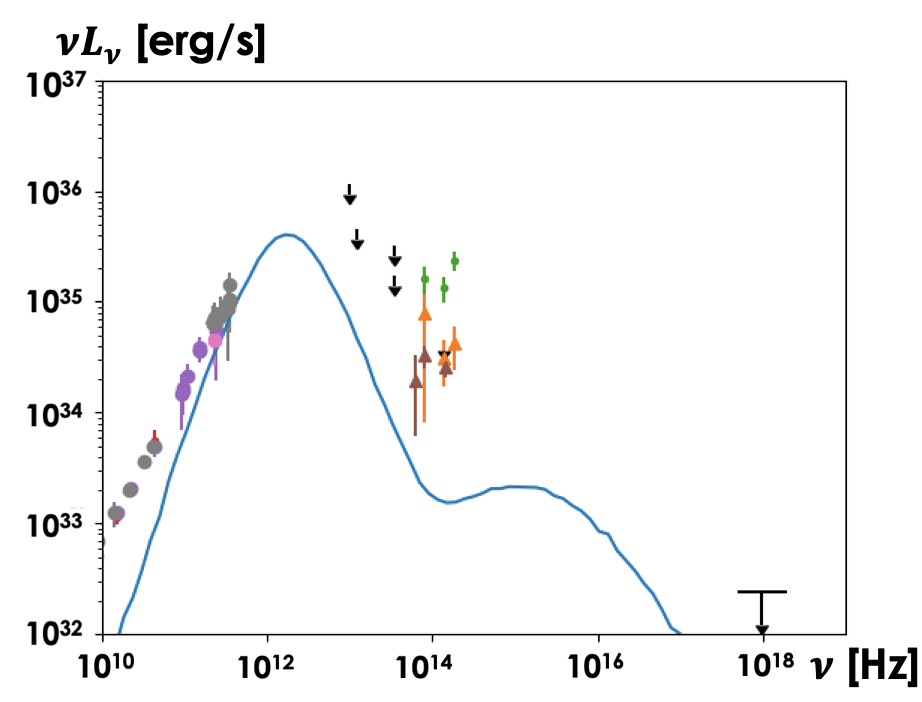}
\end{align} 
\begin{align}\nonumber 
  & \hspace{1.5cm}  \includegraphics[height=150pt,width=180pt,trim = 6mm 1mm 0mm 1mm]{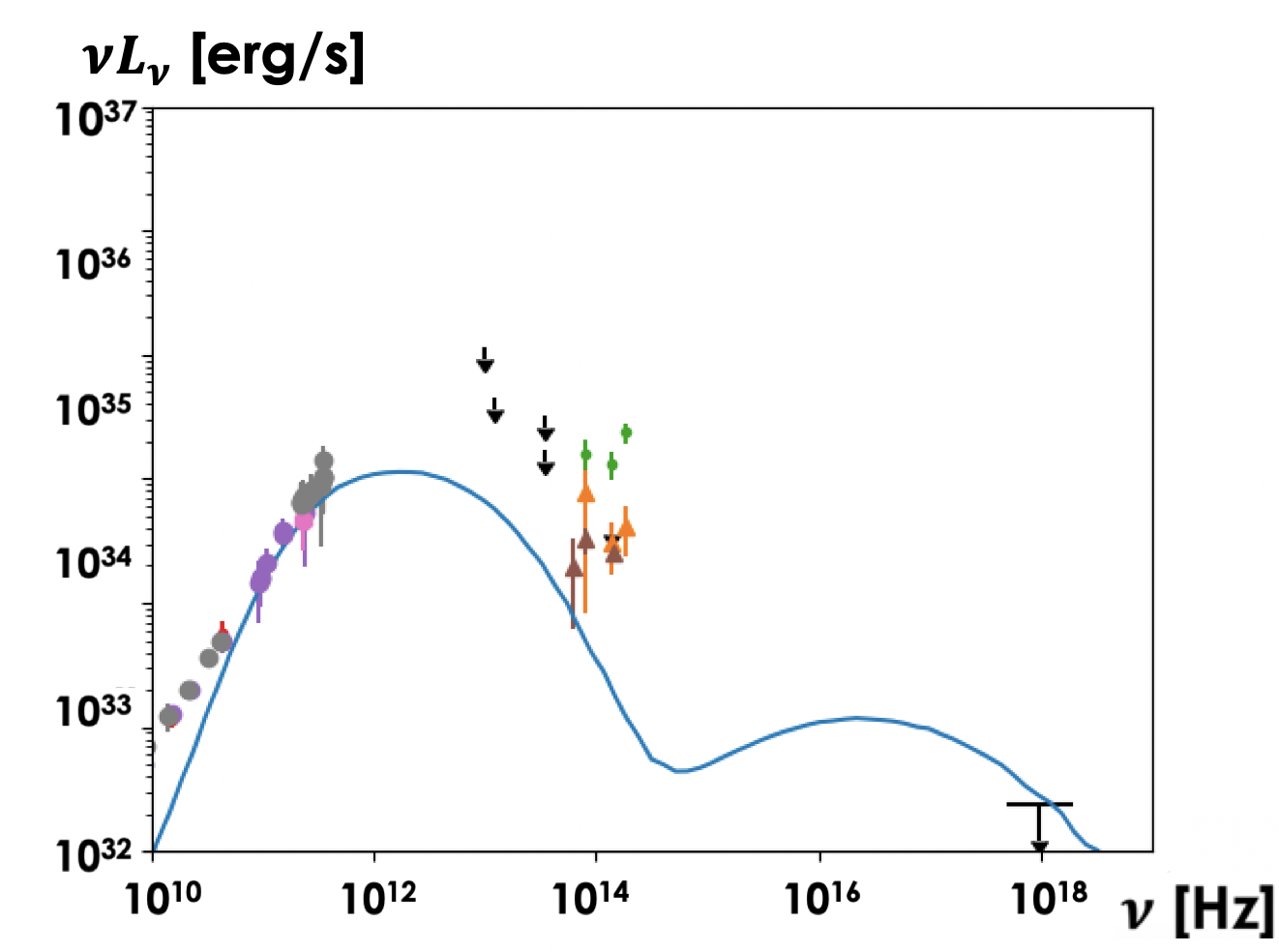
  }& 
\end{align} \caption{ 
Spectra generated from $(f,\beta_c)=(0.1,1)$ (Top Panel) and $(0.5,1)$ (Bottom Panel) Critical Beta Electron Temperature Models. 
}\label{ElectronEvolAndfAndBeta_cEqPt1And1AndfAndBeta_cEqPt5And1}
\end{figure}

Electron Temperature Model spectra generated with parameter values $(f,\beta_c)=(0.1,1)$ and $(0.5,1)$ are shown in Fig. \ref{ElectronEvolAndfAndBeta_cEqPt1And1AndfAndBeta_cEqPt5And1}. Increasing the overall electron temperature prefactor $f$ at constant $\beta_c$ is seen to increase the width of the synchrotron peak-- reflecting a greater range of emitting temperatures. For $(f,\beta_c)=(0.5,1)$, the model fits data points over a 
broad frequency range: from microwaves to infrared to X-rays, with possible improvement in the NIR with the addition of nonthermal electrons (and slight reduction in $f,\beta_c$ if this introduces an X-ray excess). From our parsimonious set of two assumptions regarding high and low $\beta$ electron temperature behavior, the simple parametrized Electron Temperature Model performs comparably to the full electron evolution calculation. In fact, for $(f,\beta_c)=(0.5,1)$, the Critical Beta  Electron Temperature Model 
is in better agreement with observations 
at the high-frequency end.

\subsection{Constant Electron Beta Model}

\subsubsection{Constant $\beta_{e}$  Model Images}

\begin{figure}
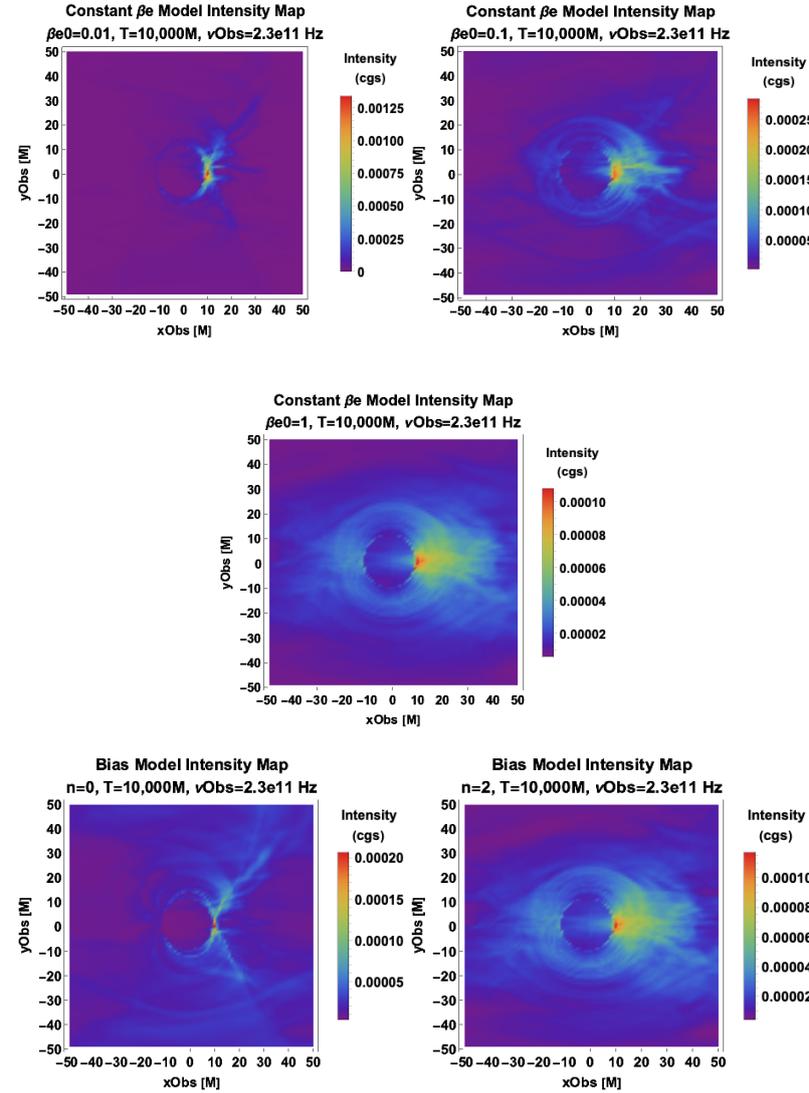
\nonumber
\begin{align}
  \hspace{-1.5mm} & \includegraphics[height=125pt,width=115pt,trim = 35mm 1mm 0mm 1mm]{PlotBetae0EqPt01ModelV2aM_unitgTotalfluxJy1Pt90238e19and2Pt40437T10000
  .png} & \includegraphics[height=125pt,width=115pt,trim = 10mm 1mm 30mm 1mm]{PlotBetae0EqPt1ModelV2aM_unitgTotalfluxJy4Pt08628e182Pt37581T10000
  .png}& 
\end{align}
\begin{align}\nonumber
  &\hspace{2.5cm} \includegraphics[height=125pt,width=115pt,trim = 35mm 1mm 0mm 1mm]{PlotBetae0Eq1ModelV2aM_unitgTotalfluxJy2Pt51415e182Pt4266T10000
  .png} & 
\end{align}
\begin{align}
  \hspace{-1.5mm} & \includegraphics[height=125pt,width=115pt,trim = 35mm 1mm 0mm 1mm]{PlotnEq0ModelV2aM_unitgTotalfluxJy9Pt18101e18and2Pt43161T10000
  .png} & \includegraphics[height=125pt,width=115pt,trim = 10mm 1mm 30mm 1mm]{PlotnEq2ModelV2aM_unitgTotalfluxJy2Pt95599e182Pt38875T10000
  .png} & 
\end{align}
\caption{Constant $\beta_e$ Model image for $\beta{e0}=0.01$ (Top Left) $\beta{e0}=0.1$ (Top Right) and $\beta_{e0}=1$ (Center), along with Bias Model images for $n=0$ (Bottom Left) and $n=2$ (Bottom Right).
}\label{EquipartitionInspiredModelImages}
\end{figure}

Constant $\beta_e$ Model images are presented for $\beta_{e0}=0.01,0.1$ and 1 in the top and middle panels of Fig. \ref{EquipartitionInspiredModelImages}. 
The images are comprised of a thick torus of lensed disk emission starting at the innermost photon rings and surrounded by dimmer filaments. Decreasing the parameter $\beta_{e0}$ decreases the electron temperature and leads to thinner torii of disk emission 
and increasing outflow-to-inflow ratio of filamentary emission. In this limit as well, we find increasing asymmetry as the photon rings emission wane into a narrow crescent on one side. 
The increased asymmetry is quantified in Appendix \ref{StatisticalAnalysis:Moments} Table \ref{tab:ImageMoments230GHz}; the centroid of the image for $\beta_{e0}=0.01$ has the furthest lateral displacement for all our models.

\subsubsection{Constant $\beta_{e}$ Model Spectra}
  
\begin{figure}
\begin{align}\nonumber
  & \hspace{4mm} \includegraphics[height=120pt,width=130pt,trim = 15mm 1mm 0mm 1mm
  ]{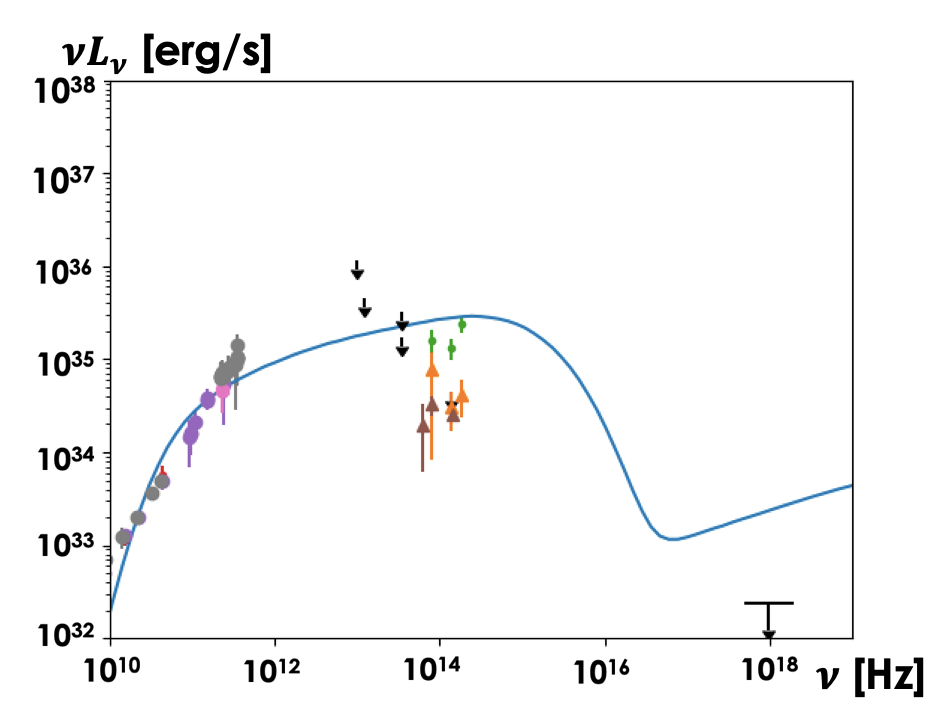
  } & \hspace{0.5mm}\includegraphics[height=120pt,width=130pt,trim = 15mm 1mm 0mm 1mm
  ]{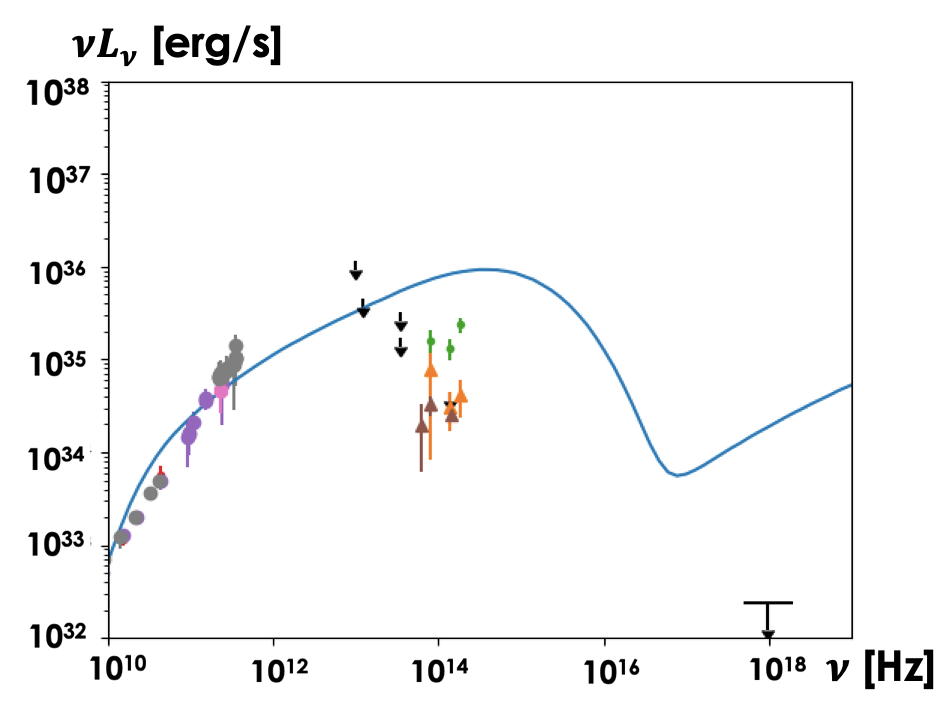
  } &  
\end{align}
\begin{align}\nonumber
  & \hspace{4mm}    \includegraphics[height=120pt,width=130pt,trim = 15mm 1mm 0mm 1mm]{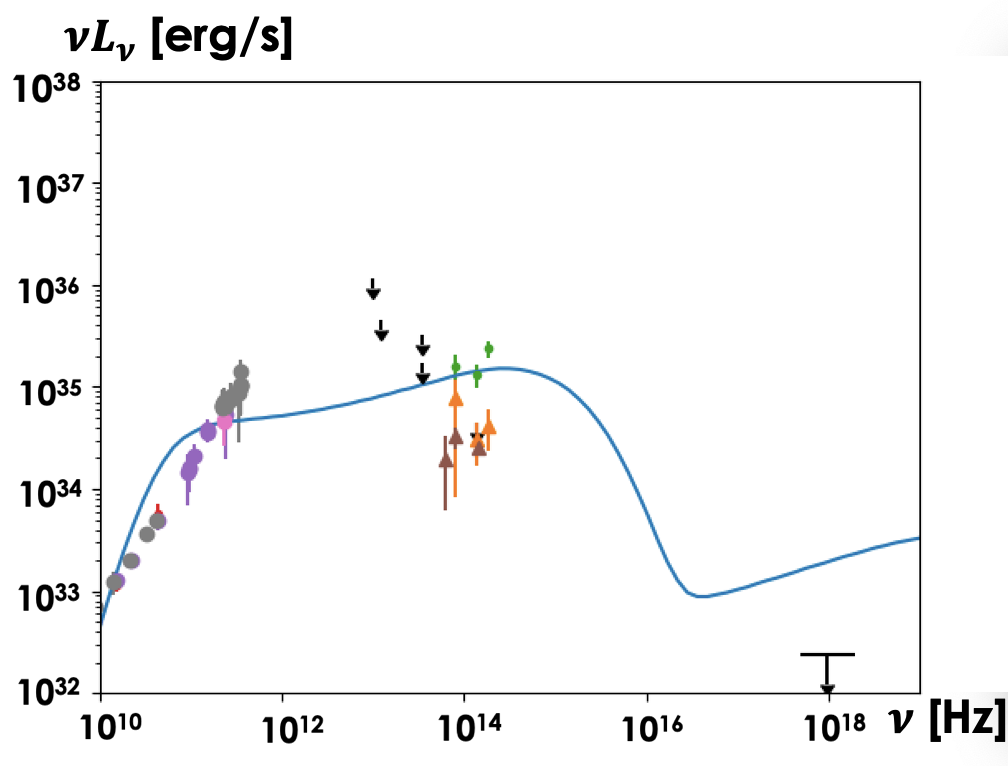
  } & \hspace{0.5mm} \includegraphics[height=120pt,width=130pt,trim = 15mm 1mm 0mm 1mm]{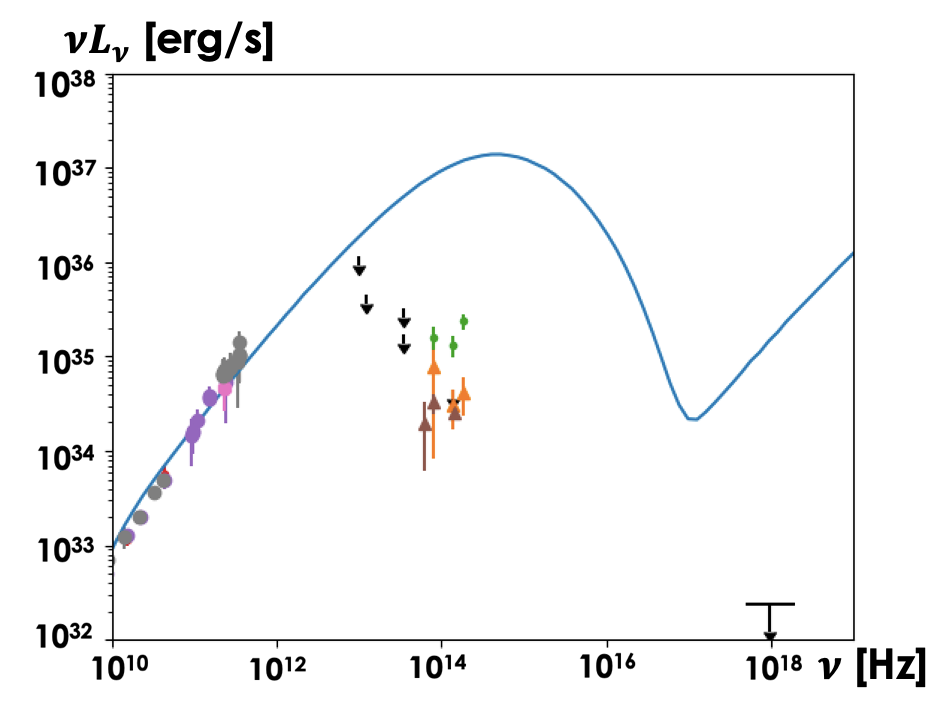
  } &  
\end{align} 
\caption{Observed spectra (dots) compared with synthetic model spectra (curves) generated from 
$\beta_{e0}=0.01$ Model
(Top Left),
$\beta_{e0}=0.1$ Model (Top Right)
, 
$
n=0$ Bias Model (Bottom Left) 
and 
$
n=2$ Bias Model
(Bottom Right).
}\label{SpectralComparisonWithObservationsr<30}
\end{figure}

Constant $\beta_e$ Model spectra for $\beta_{e0}=0.01$ and 0.1 are shown in Fig. \ref{SpectralComparisonWithObservationsr<30}  
excluding regions $r>30M$ that are not in equilibrium. These spectra generally reproduce the low frequency slope well, though flatten near the infrared bump-- particularly for lower values of the constant $\beta_e$. 
 The radio spectra can be explained in the context of the Blandford-K\"onigl model \citep{BK1979}, which demonstrates that helical magnetic fields and constant  $\beta_e$ in outflows that are optically thick to synchrotron emission produce radio spectra that are flat in $ L_\nu$, consistent with the low frequency emission in Sgr A*.  The spectral slope rises in the X-ray, as outflow/jet emission overproduces the high-frequency spectrum.

\subsection{Magnetic Bias Model}

\subsubsection{Bias Model Images}

From Eq. \ref{BiasModel}, it is manifest that the simplest ($n=0$) Bias Model has constant relativistic electron gas pressure throughout the simulation, in contrast to its expected decrease along the outflow for higher values of $n$. The image in the bottom left panel of Fig. \ref{EquipartitionInspiredModelImages}  
for $n=0$ shows extended emission tracing a funnel shape in the jet/outflow region. For $n=2$,  
emission becomes dominated by 
a thick photon torus lensed orthogonally to the accretion disk.

For Bias Model images, decreasing $n$ leads to more extended outflow contributions to emission.
The Bias and Constant $\beta_e$ Models have the most drastic variation of image intensity and shape over the observer plane for the models considered in this work.

\subsubsection{Bias Model Spectra}
 
Spectra for 
Bias Models with $n=0$ and 2 
are shown in Fig. \ref{SpectralComparisonWithObservationsr<30}. 
The $n = 2$ model dramatically overproduces the emission at IR-X-rays. Both the $n = 0$ and $n = 2$ models do a reasonable job of explaining the low frequency radio emission, which is not surprising since these models are generalizations of the constant $\beta_e$ model known to explain the radio emission in optically thick AGN jets.
 
 

\subsection{EHT and GRAVITY}

\citet{Dexter2010} fit Sgr A* thermal synchrotron emission models parameterized by mass accretion rate, orientation angle, spin and electron-to-proton temperature ratio to mm-VLBI and spectral data \citep{Marrone2006SubmillimeterPO}, favoring a wide $50^\circ$ viewing angle. Our view of the Galactic Center has evolved, as 
GRAVITY has provided indications from infrared observations that the Galactic Center inner disk (between 6$M$ and 10$M$) appears more face on \citep{Gillessen2017}. 

\begin{figure}  
\begin{align}\nonumber
  & \hspace{-2.5mm} \includegraphics[height=125pt,width=115pt,trim = 15mm 1mm 0mm 1mm
  ]{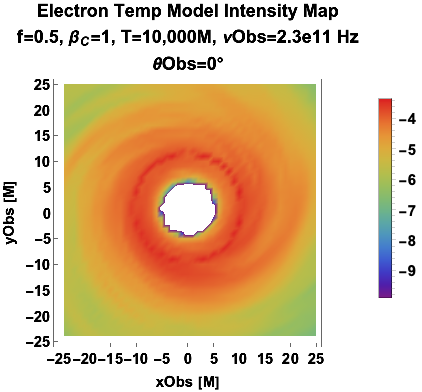} & \includegraphics[height=125pt,width=115pt,trim = 6mm 1mm 20mm 1mm]{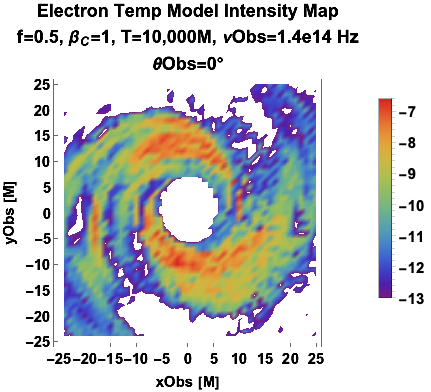} &   
\end{align}
\begin{align}\nonumber
  & \hspace{-2.5mm} \includegraphics[height=125pt,width=115pt,trim = 15mm 1mm 0mm 1mm]{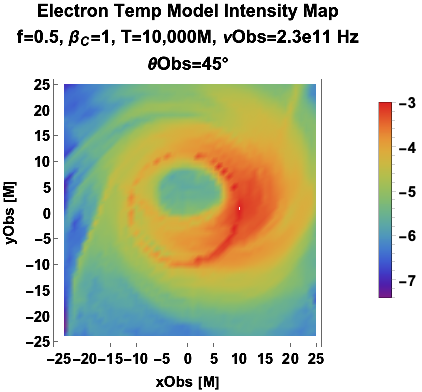} & \includegraphics[height=125pt,width=115pt,trim = 6mm 1mm 20mm 1mm]{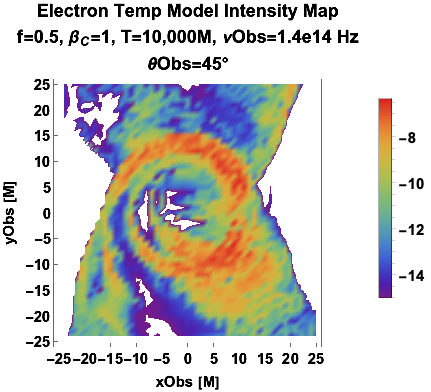} & 
\end{align}
\begin{align}\nonumber
  & \hspace{-2.5mm}  \includegraphics[height=125pt,width=115pt,trim = 15mm 1mm 0mm 1mm]{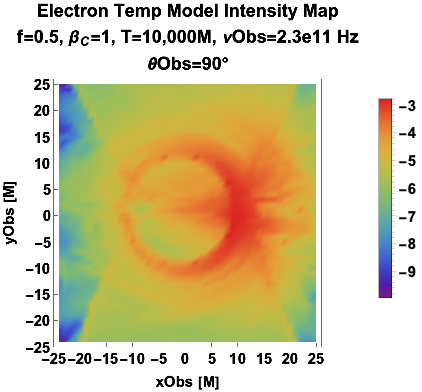} & \includegraphics[height=125pt,width=115pt,trim = 6mm 1mm 20mm 1mm]{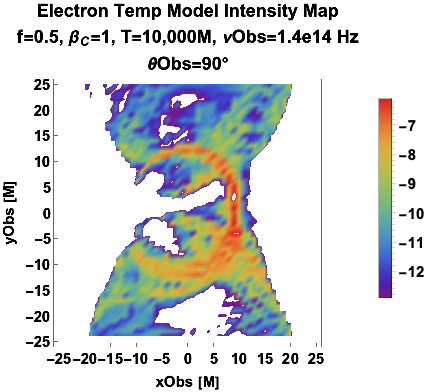} & 
\end{align}
\caption{EHT scale (230 GHz, Left) versus GRAVITY scale (140 THz, Right) logplot images of the favored
$(f,\beta_{c})=(0.5,1)$ Model. The observer angle varies from $\theta_\mathrm{Obs}\approx0$ to $\frac{\pi}{4}$ to $\frac{\pi}{2}$ from Top to Bottom. These image maps are shown on a log scale in order to accentuate features in the intensity profiles.  
}\label{EHTvsGravityComparison}
\end{figure}

We now compare synthetic EHT-scale (230 GHz) images against GRAVITY-scale ($2.2\ \mu$m, or 1.4 THz) images for 
a particular model, $(f,\beta_{c})=(0.5,1)$, in Fig. \ref{EHTvsGravityComparison}, varying the viewing angle from face-on to edge-on (note, we display these on a common log scale to accentuate features). The face-on disk has smoother variation of intensity with radius at 230 GHz, as it slowly goes from spiral edges to circular at $r\sim 10M$ to the innermost stable circular orbit.  At 2.2 $\mu$m, the disk appears spiral throughout, punctuated with distinct bright spots. At 45$^\circ$ inclination, the brightest feature at both frequencies is a circular ring with $r\lesssim 15M$. The edge-on view exhibits greater disk asymmetry due to Doppler brightening at 230 GHz, and has a more prominent outflow with more apparent substructure 
at 2.2 $\mu$m.  
 
\begin{figure}  
\begin{align}\nonumber
  & \hspace{1mm} \includegraphics[height=125pt,width=115pt,trim = 15mm 1mm 0mm 1mm
  ]{Log10PlotfEqPt5BetaTotCritEq1Zoom_M_unit__g__Total_flux__Jy____3_94188e19_4_01178em4__T10000_nuEq1Pt4e14Hz_Rangem13Tom6_ThetaObsEq0.png} & \includegraphics[height=125pt,width=115pt,trim = 6mm 1mm 20mm 1mm]{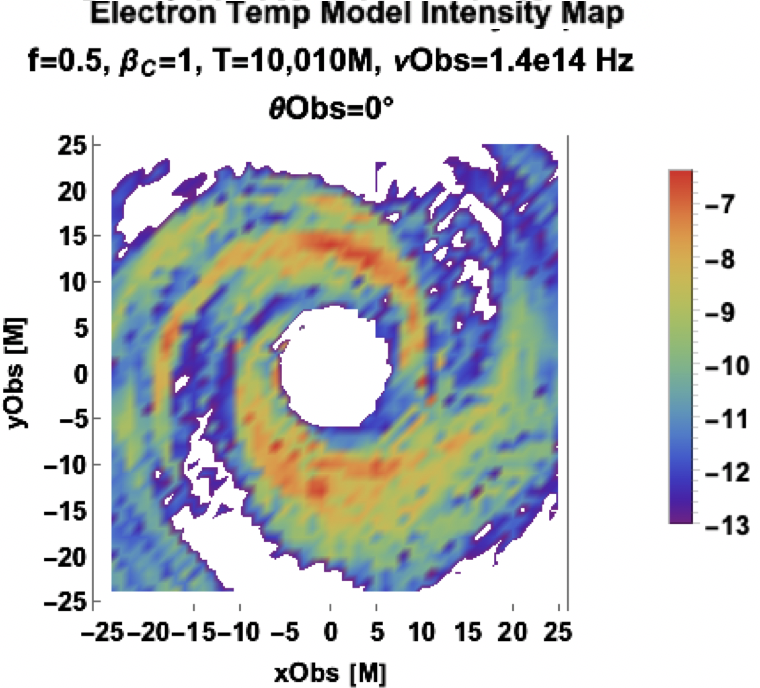} &   
\end{align}
\begin{align}\nonumber
  & \hspace{1mm} \includegraphics[height=125pt,width=115pt,trim = 15mm 1mm 0mm 1mm]{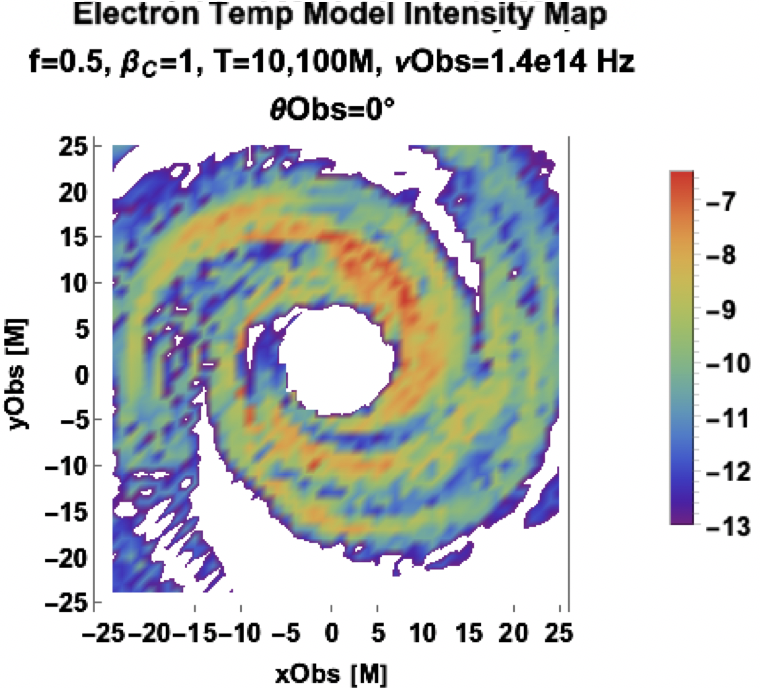} & \includegraphics[height=125pt,width=115pt,trim = 6mm 1mm 20mm 1mm]{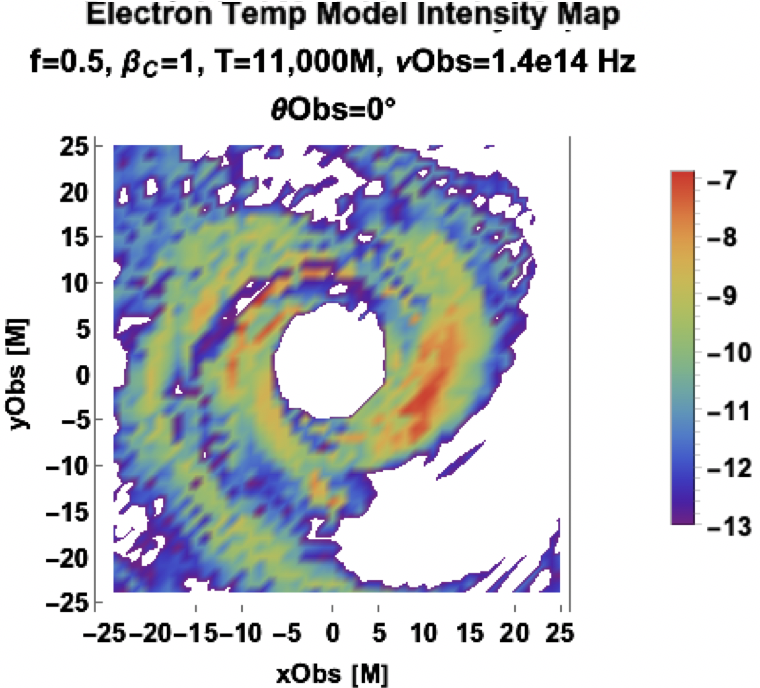} & 
\end{align}
\caption{A sequence of face-on $(f,\beta_{c})=(0.5,1)$ Model infrared images at $T=10,000M,10,010M,10,100M$ and 11,000$M$. Variation can be observed after a hundred $M$, but the basic spiral morphology with bright spots is maintained over at least a thousand $M$. 
}\label{IRFlares}
\end{figure}

We see the dynamical behavior of  the 
$(f,\beta_c)=(0.5,1)$ model 
at IR frequency in Fig. \ref{IRFlares}, where we show face-on images at several different times. The bright spots appear to rotate on time scales between $\Delta T= 100M$ and $1,000M$ where $M=20$s. The image symmetry makes the centroid motion localized to small gyrations centered on the black hole for this model.
 
 \subsection{SANE vs. SEMI-MAD vs. MAD Simulations}

\begin{figure}
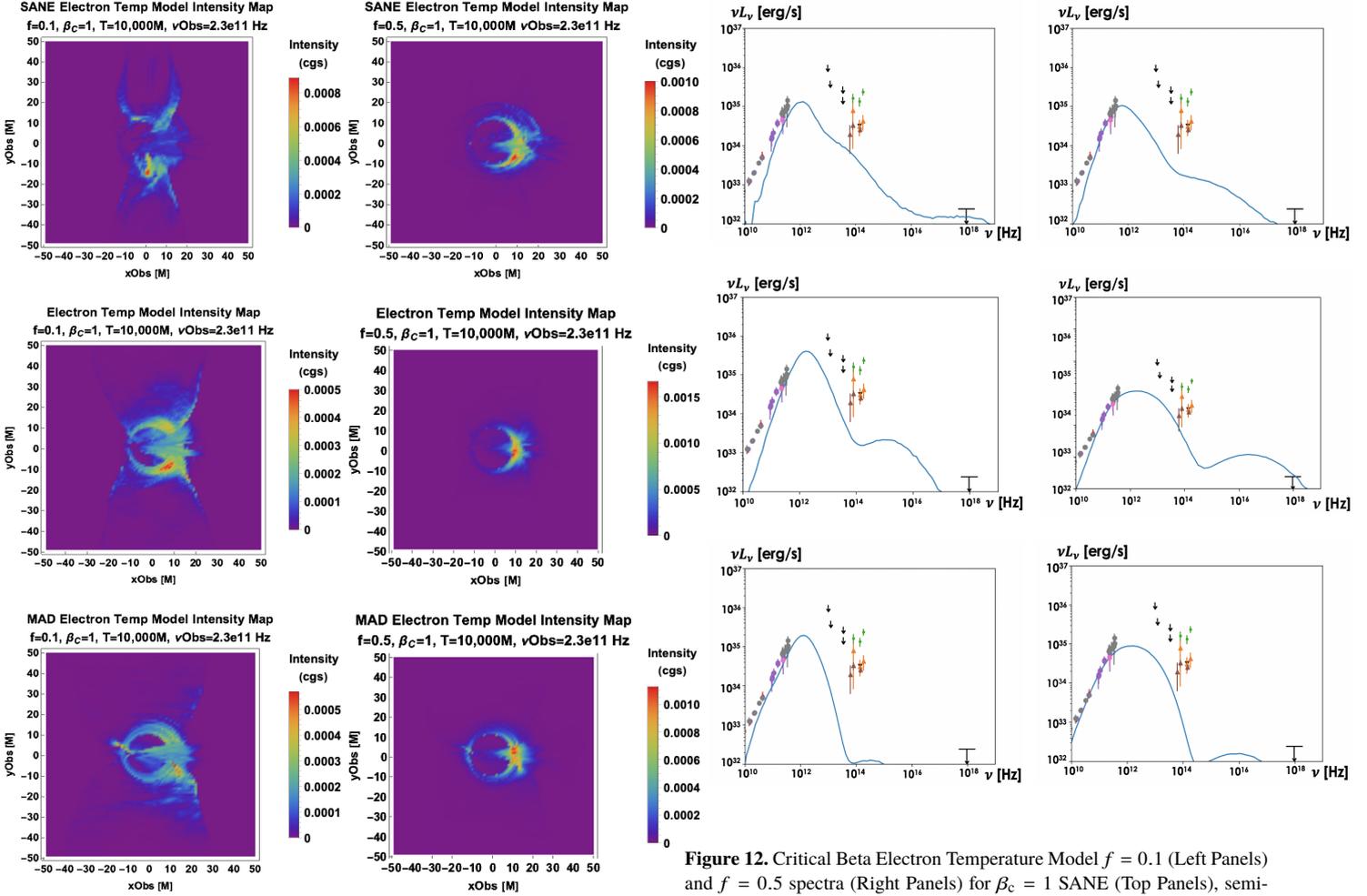
\nonumber
\begin{align}\nonumber
 \hspace{5mm} & \includegraphics[height=115pt,width=110pt,trim = 35mm 1mm 0mm 1mm]{PlotfEqPt1BetaCritEq1SANEModelV1M_unitgTotalfluxJy4Pt22646e20and2Pt39825T10000
.png} & \includegraphics[height=115pt,width=110pt,trim = 10mm 1mm 30mm 1mm]{PlotfEqPt5BetaCritEq1SANEModelV1M_unitgTotalfluxJy1Pt85173e19and2Pt39114T10000
  .png} &  
\end{align}
\begin{align}\nonumber
 \hspace{5mm}  & \includegraphics[height=115pt,width=110pt,trim = 35mm 1mm 0mm 1mm]{PlotfEqPt1BetaTotCritEq1V1fBetaTotCrit0Pt1and1M_unitgTotalfluxJy5Pt94859e20and2Pt40018T10000
  .png} & \includegraphics[height=115pt,width=110pt,trim = 10mm 1mm 30mm 1mm]{PlotfEqPt5BetaTotCritEq1V1fBetaTotCrit0Pt5and1M_unitgTotalfluxJy3Pt94188e19and2Pt3596T10000
  .png} & 
\end{align}
\begin{align}
  \hspace{5mm} & \includegraphics[height=115pt,width=110pt,trim = 35mm 1mm 0mm 1mm]{PlotfEqPt1BetaCritEq1MADModelV1M_unitgTotalfluxJy1Pt3886e18and2Pt40796T10000
  .png} & \includegraphics[height=115pt,width=110pt,trim = 10mm 1mm 30mm 1mm]{PlotfEqPt5BetaCritEq1MADModelV1M_unitgTotalfluxJy1Pt06979e17and2Pt43971T10000
  .png} & 
\end{align}
\caption{Critical Beta Electron Temperature Model  for $f=0.1$ (Left Panels) and $f=0.5$ images (Right Panels) for $\beta_\mathrm{c}=1$ SANE (Top Panels), semi-MAD (Middle Panels) and MAD simulations (Bottom Panels).}\label{fEqPt1orPt5ElectronTemperatureModelSANEsemiandMAD}
\end{figure}

 Standard and normal evolution (SANE), magnetically arrested disk (MAD) and semi-MAD simulations represent quite distinct forms of evolution of magnetized accretion flows. The SANE case has the lowest magnetic flux; the MAD case admits the lowest amount of mass accreted by the black hole, as in Table \ref{MUnit230GHz}. Compared to images ray traced from the fiducial semi-MAD simulation, the SANE and MAD images vary in size and asymmetry in a similar manner with changing parameters in our models. For example, Fig. \ref{fEqPt1orPt5ElectronTemperatureModelSANEsemiandMAD} shows that increasing Critical Beta Electron Temperature Model parameter $f$ from 0.1 to 0.5 takes a relatively symmetric photon ring/torus  and extended outflow structure to a compact, asymmetric ring for all three simulations. The same trend holds for increasing $\beta_c$ from 0.01 to 1. For the $f=0.5$ models in all three simulations, a disk feature is visible emanating from the bright spot and extending along the projected equatorial plane. It is also noteworthy that the best $10^{11}$ Hz $<\nu_\mathrm{Obs}<$ $10^{19}$ Hz spectrum across simulations (cf. Fig. \ref{SpectrafEqPt1orPt5ElectronTemperatureModelSANEsemiandMAD}) corresponds to the $(f,\beta_c)=(0.5,1)$ model in the semi-MAD simulation; however, the MAD simulation for this model generates the best low frequency fit. We do add the caveat that with the addition of non-thermal particles, we expect the MAD images to have increased NIR and X-ray contributions. 
 As for other models, decreasing Constant Electron Beta Model Parameter $\beta_{e0}$ from 1 to 0.01 changes images from thick photon torii 
 to thin rings + outflows. Decreasing Magnetic Bias Model parameter $n$ from 2 to 0 accentuates the outflow in all 3 simulations, with the key difference being greatest collimation in the SANE simulation. For all simulations and models, inclusion of non-thermal particles could readily increase the IR and X-ray emission.

\begin{figure}\nonumber
\begin{align}\nonumber
  &\hspace{5mm} \includegraphics[height=100pt,width=125pt,trim = 6mm 1mm 0mm 1mm]{SpectrumfEqPt1BetaCEq1Munit4Pt50112e20Flux2Pt39665SANEwAxesLabels
.png} & \includegraphics[height=100pt,width=125pt,trim = 6mm 1mm 0mm 1mm]{SpectrumfEqPt5BetaCEq1Munit2Pt55178e19Flux2Pt42829SANEwAxesLabels
  .png} & 
\end{align}
\begin{align}\nonumber
  & \hspace{5mm}\includegraphics[height=100pt,width=125pt,trim = 6mm 1mm 0mm 1mm]{SpectrumfEqPt1BetaTotCritEq1MUnit5Pt94859e20gFlux2Pt40018JyRange10e32To10e37wAxesLabel
  .png} & \includegraphics[height=100pt,width=125pt,trim = 6mm 1mm 0mm 1mm]{SpectrumfEqPt5BetaTotCritEq1MUnit3Pt94188e19gFlux2Pt3596JyRange10e32To10e37wAxesLabel
  .png} & 
\end{align}
\begin{align}
  &\hspace{5mm}\includegraphics[height=100pt,width=125pt,trim = 6mm 1mm 0mm 1mm]{SpectrumfEqPt1BetaCEq1Munit1Pt67729e18Flux2Pt39972MADwAxesLabels
  .png} &\includegraphics[height=100pt,width=125pt,trim = 6mm 1mm 0mm 1mm]{SpectrumfEqPt5BetaCEq1Munit1Pt40117e17Flux2Pt42936MADwAxesLabels
  .png} &  
\end{align}
\caption{Critical Beta Electron Temperature Model $f=0.1$  (Left Panels) and $f=0.5$ spectra (Right Panels) for $\beta_\mathrm{c}=1$ SANE (Top Panels), semi-MAD (Middle Panels) and MAD simulations (Bottom Panels). }\label{SpectrafEqPt1orPt5ElectronTemperatureModelSANEsemiandMAD}
\end{figure}



\subsection{mm-/NIR-
Variability}

\begin{figure}\nonumber
\begin{align}
 \hspace{1.3cm} & \includegraphics[height=170pt,width=200pt,trim = 35mm 1mm 0mm 1mm]{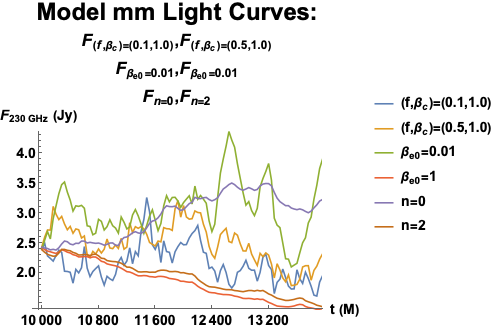} 
\end{align} 
\begin{align}\nonumber
 \hspace{0.3cm} &
\includegraphics[height=170pt,width=200pt,trim = 10mm 1mm 30mm 1mm]{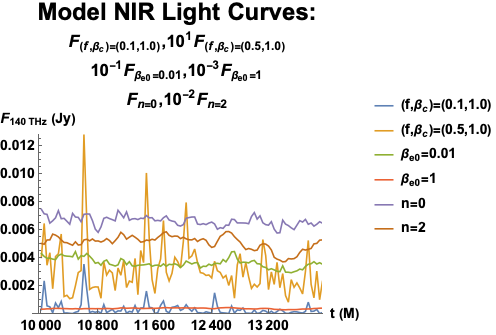} 
\end{align}
\caption{Light curves at 230 GHz (Top)  and 140 THz (Bottom) for parametric models with $(f,\beta_c)=(0.1,1)$, $(f,\beta_c)=(0.5,1)$,  $\beta_{e0}=0.01$, $\beta_{e0}=1$, $n=0$  and $n=2$.
}\label{ModelLightCurves}
\end{figure}

In Fig. \ref{ModelLightCurves}, we display mm- and NIR light curves for select models.  The models exhibit different variability timescales, e.g., doubling times, over the 4,000$M$ ($\approx24$h) interval depicted. The $\beta_{e0}=1$ and $n=2$ models exhibit the mildest variation of a few percent within an hour ($\approx 160M$).  These models are immediately precluded if we require at least minute-scale variability. The $n=0$ model is slightly more variable, with slightly larger amplitude local minima and maxima. The $(f,\beta_c)=(0.5,1)$ model is the most rapidly varying, followed by $(f,\beta_c)=(0.1,1)$. The greatest amplitude flare-like feature in the radio is the two-fold peak-to-trough variation for $12,700M\lesssim t_\mathrm{Obs} \lesssim  13,500M$ in the $\beta_{e0}=0.01$ model light curve. 

The NIR luminosity differs by orders of magnitude among the models in Fig. \ref{ModelLightCurves}.  The models with the greatest overall flux are for $n=2$ and  $\beta_{e0}=1$. The lowest flux model, $(f,\beta_c)=(0.5,1)$, exhibits the fastest peak-to-trough variability and the largest percentage swings.  But despite these large relative excursions from the mean flux, the low absolute amplitude NIR luminosity suggests that this model requires a non-thermal particle contribution. 

\subsubsection{Time Averaging}\label{TimeAveraging}
We have primarily compared models at a particular time $T=10,000M$ for which the simulation flow is statistically in steady state.  However, given the $20$s light crossing time of Sgr A*'s gravitational radius, variability challenges EHT's temporal imaging capacity for Sgr A*-- and all but the most massive black holes.  
Fig. \ref{TimeAveragedImage} shows an image averaged over $1000 \, M$ (6 hr for Sgr A*) for our $(f,\beta_c) = (0.5,1)$ model.  The result is very similar to the instantaneous image in Fig. \ref{ElectronTemperatureModelImages}.   Examining time evolution of the image in detail, we find that there is image variability on roughly hourly timescales, with some bright features moving around.   The overall image morphology is, however, relatively stable.

 
\section{Comparison of Models}

We now synthesize our results to assess which models are favored by observations, starting with our fiducial semi-MAD simulation.
A summary of how the different models fare against observational constraints is given in Table \ref{PassFail}.

\subsection{Comparison of Images}

In the images for the Electron Evolution Model with Turbulent Heating, the emission is concentrated in an asymmetric photon ring, with some some contribution from outflow at small radii. The Critical Beta  Electron Temperature Model images have emission smeared out broadly over inflow and outflow 
at low $f$ or $\beta_c$, and approach a compact, asymmetric photon ring for $(f,\beta_c)\to(0.5,1)$, with even less outflow contribution relative to the Electron Evolution Model. The Constant $\beta_e$ and Bias Models have drastically varying morphology over the parameter space scanned-- from long, extended outflow filaments for the low $\beta_{e0}$ and low $n$ models, to thick photon torii 
in the high $\beta_{e0}$ and high $n$ limits.

The relative asymmetry of model images can be compared using image moments in Appendix \ref{StatisticalAnalysis:Moments} Table \ref{tab:ImageMoments230GHz}, where models in which most of the flux density emanates from compact regions regions such as $\beta_{e0}=0.01$ or $(f,\beta_c)=(0.5,1)$ are shown to have centroids shifted furthest to the right in our fiducial simulation. We make 
further quantitative comparisons of the images
by ascribing emitting region sizes to our intensity maps.

\subsubsection{Emitting Region}

\begin{table*}
	\centering
	\caption{Elliptical Gaussian semi-axes $\theta_\mathrm{min}$ and $\theta_\mathrm{maj}$ generated from second order central image moments and image covariance matrices for our models for SANE, MAD and semi-MAD simulations at the fiducial 230 GHz frequency and 90$^\circ$ (edge-on) viewing angle.  The anisotropic observational size constraints  are $5.44M<\theta_\mathrm{min}<10.68M$ and  $9.87M<\theta_\mathrm{maj}<12.69M$ \citep{Johnson_2018}.
    }
	\label{tab:Gtr10AndGtr20PctAreas}
	\begin{tabular}{lccccccr} 
		\hline
		Model & 
        && 
        $\frac{\theta_\mathrm{min\ }}{M}=\sqrt{(8\ln{2})\lambda_\mathrm{min}}$ && 
        & 
        $\frac{\theta_\mathrm{maj\ }}{M}=\sqrt{(8\ln{2})\lambda_\mathrm{max}}$ &  \\
        &&&&&&\\
     & 
     &SANE&  semi-MAD  &MAD& 
     SANE& semi-MAD & MAD  \\
     		\hline
	 Electron Evolution with Turbulent Heating  
          & 
          &
          \textbf{8.2}
          & 
          \textbf{9.0}
          &
          11.4
          &
          18.2
          & 
          15.3
          &
          15.1
          \\
		\hline
		Critical Beta Electron Temperature ($f,\beta_c$) &    & & & \\
        ($0.1,0.01$)& 
        &
        \textbf{9.4}
        &   
        24.7
        &
        38.9
        & 
        28.8
        & 
        48.4
        &
        73.1
        \\
       ($0.1,0.1$)&  
       &
         \textbf{8.8}
       &  
       21.8
       &
       18.5
       & 
       24.7
       &  
       39.7
       &
       34.2
       \\
       ($0.1,1.0$)&  
       & 
       \textbf{8.8}
       &   
       20.8
       & 
       15.0
       & 
       22.3
       &  
       33.6
       &
       23.7
       \\
       ($0.5,0.01$) & 
       &
       \textbf{9.4}
       &   
       33.6
       & 
       25.0
       & 
       29.2
       &  
       43.7
       &
       55.9
       \\
       ($0.5,0.1$)&  
       &
       \textbf{8.7}
       & 
       16.8
       &
       19.1
       & 
       22.8
       & 
       24.4
       &
       34.2
       \\
       ($0.5,1.0$)&  
       &
       \textbf{9.3}
       & 
       15.3
       &
       12.5
       & 
       \textbf{11.7}
       & 
       18.5
       &
       14.3
       \\ 
        \hline
		Constant Electron Beta ($\beta_{e0}$)&  &  & & &\\
         0.01   &  
         &
         14.9
         &   
         22.7
         &
         22.5
         & 
         45.7
         & 
         44.2
         &
         44.2
         \\
         0.1   & 
         &
         19.0
         &  
         45.4
         &
         47.2
         & 
         26.9
         & 
         46.7
         &
         53.4
         \\
         1.0   & 
         &
         37.2
         &  
         45.9
         &
         57.7
         & 
         39.7
         & 
         56.0
         &
         58.7
         \\
        \hline
		 Magnetic Bias ($n$)
         &  &  & & & \\
		0 &  
        &
        43.7
        &  
        58.6
        &
        30.2
        & 
        75.1
        & 
        66.9
        &
        75.3
        \\
        1 &  
        &
        37.2
         &  
         45.9
         &
         57.7
         & 
         39.7
         & 
         56.0
         &
         34.2
        \\
        2 & 
        &
        30.7
        & 
        43.1
        &
        60.5
        & 
        47.3
        & 
        52.9
        &
        62.8
        \\ 
        \hline
	\end{tabular}\label{AnisotropicGaussianSize}
\end{table*}


We adopt the image covariance approach that \citet{Johnson_2018} used to derive the $9.87M<\theta_\mathrm{maj,Obs}
<12.69M$ elliptical Gaussian size constraint for 230 GHz Sgr A* EHT observations to formulate characteristic emitting region sizes for our synthetic images. The intensity can be expressed as a moment generating function

\addtocounter{equation}{7}

\begin{equation}
\tilde{I}(\mathbf{u})=\int d^2\mathbf{x} I(\mathbf{x})e^{-2\pi i \mathbf{u}\cdot\mathbf{x}}    
\end{equation}
The image covariance matrix is 
   \begin{align}\label{ImageCovariance}
M =   \begin{pmatrix} \frac{M_{20}}{M_{00}}-\left(\frac{M_{10}}{M_{00}}\right)^2 & \frac{M_{11}}{M_{00}}-\frac{M_{10}M_{01}}{M_{00}^2} \\
  \frac{M_{11}}{M_{00}}-\frac{M_{10}M_{01}}{M_{00}^2} & \frac{M_{02}}{M_{00}}-\left(\frac{M_{01}}{M_{00}}\right)^2  \\
  \end{pmatrix}.
\end{align}
where 
\begin{equation}
M_{n_1,n_2}=\sum_{i,j}x_i^{n_1}y_j^{n_2}I(x_i,y_j)
\end{equation}
The equivalent Gaussian FWHM for images in Observer Plane coordinates is
\begin{align}\nonumber
(\theta_\mathrm{maj})_\mathrm{Obs} = \sqrt{-\frac{2\ln 2}{\pi^2\int d^2\mathbf{x} I(\mathbf{x})}}\times \hspace{4.5cm}\\ \nonumber
\sqrt{\left((\hat{u}_\mathrm{maj})_1^2\frac{\partial^2\tilde{I}}{\partial u_1^2}+2(\hat{u}_\mathrm{maj})_1(\hat{u}_\mathrm{maj})_2\frac{\partial^2\tilde{I}}{\partial u_1 \partial u_2}+ (\hat{u}_\mathrm{maj})_2^2\frac{\partial^2\tilde{I}}{\partial u_2^2}\right)}
\end{align}

\begin{equation}
=2\sqrt{2\ln 2\frac{(\hat{u}_\mathrm{maj})_1^2M_{20}+2(\hat{u}_\mathrm{maj})_1(\hat{u}_\mathrm{maj})_2M_{11}+(\hat{u}_\mathrm{maj})_2^2M_{02}}{M_{00}}}
\end{equation}
where $\hat{u}_\mathrm{maj}$ is the eigenvector of the covariance matrix corresponding to greatest eigenvalue $\lambda_\mathrm{maj}$ and $\hat{u}_\mathrm{min}$ is defined analogously. We may conceptualize $(\theta_\mathrm{maj})_\mathrm{Obs}$ as a normalized second directional derivative along the image major axis. We take our characteristic image size to be the centroid-adjusted equivalent Gaussian FWHM
\begin{equation}
\theta_\mathrm{maj} = 2\sqrt{(2\ln{2})\lambda_\mathrm{maj}}
\end{equation}

In Table \ref{AnisotropicGaussianSize
} using the elliptical Gaussian size constraint, we have the following trends in image size:
\begin{itemize}
    \item Image size decreases with $f$ or $\beta_c$
    \item Image size increases with $\beta_{e0}$ (with the exception of the SANE simulation at low $\beta_{e0}$
    \item Image size decreases with $n$ (exceptions: high $n$ SANE and MAD)
\end{itemize}
The preferred models with smallest size occur for $(0.5,1)$, particularly in the SANE simulation, which satisfies the EHT size constraint for $\theta_\mathrm{maj}$.

\begin{figure} 
\begin{align}\nonumber
  & \hspace{1.0cm} \includegraphics[height=175pt,width=200pt,trim = 6mm 1mm 0mm 1mm]{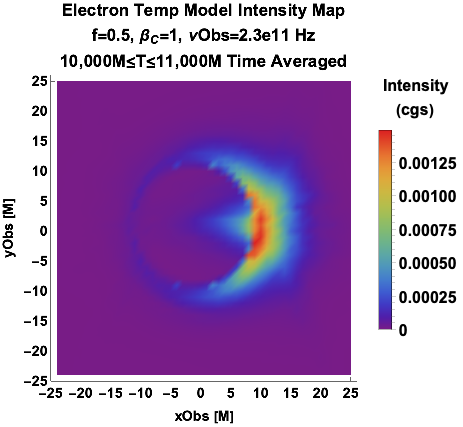}
  & 
\end{align}
\caption{Preferred model ($(f,\beta_c)=(0.5,1)$) averaged over simulation times $10,000M<T<11,000M$ at 230 GHz.}\label{TimeAveragedImage}
\end{figure}

Note that observational effects such as scattering, the point spread function and instrument-specific cadence tend to wash out image features. While we do not include a model for scattering in this work, we have time-averaged our preferred model in 
Fig. \ref{TimeAveragedImage} to reflect uncertainties due to temporal resolution. 
The time averaged image for the $\sim 6$h interval $10,000M<T<11,000M$ is similar to the instant one in Fig. \ref{ElectronTemperatureModelImages} (Lower Right), and the spectral shape does not vary appreciably even over the $\sim 12$h interval $9,000M<T<11,000M$, as discussed in Section \ref{TimeAveraging}.

\subsubsection{Image Size Frequency and Viewing Angle Dependence}

In Table \ref{tab:86GHzVs230GHzAnd90DegVs0Deg}, image sizes are compared at angles $0^\circ$ (face on) vs. $90^\circ$ (edge on) and frequencies 86 GHz (Global mm VLBI Array (GMVA)) vs. 230 GHz (EHT). For both frequencies, the edge-on image sizes tend to decrease with each of $f$ and $\beta_c$, increase in $\beta_{e0}$ and decrease in $n$; the face-on images do not exhibit monotonic behavior. Moreover, at a given frequency, the edge-on images tend to be larger than their face-on parametric counterparts for Critical $\beta$ Models (except at high $(f,\beta_c)$); face-on images are larger than edge-on images for the Constant $\beta_e$ Model (except at low $\beta_{e0}$) and Magnetic Bias Model (except at higher frequency and low $n$). Face on and at low frequency, intensity profiles appear broad and point symmetric due to the projection of co-axial outflows and disk onto the observer plane-- particularly for low $n$ and low $(f,\beta_c)$. The smallest of our flux-normalized images tend to occur for the most asymmetric models, e.g., $(f,\beta_c)=(0.5,1)$.

The images satisfying the 86 GHz size constraint occur for the Critical Beta Electron Temperature Models $(f,\beta_c)=(0.1,0.01),(0.1,0.1),(0.1,1),(0.5,0.01),(0.5,0.1)$ and $(0.5,1)$ for the SANE and MAD simulations and MAD $\beta_{e}=0.01,0.1$ and 1 Constant $\beta_{e0}$ and $n=0,1$ and 2 Bias Models viewed at $0^\circ$; and the $(f,\beta_c)=(0.5,1)$ Model for the SANE, semi-MAD and MAD simulations at 
$90^\circ.$ 
The unique model satisfying the 230 GHz size constrain is the $(f,\beta_c)=(0.5,1)$ Model in the SANE simulation viewed edge on. 

\begin{table*}
	\centering
	\caption{Elliptical Gaussian semi-axes $\theta_\mathrm{min}$ and $\theta_\mathrm{maj}$ generated from second order central image moments and image covariance matrices for our semi-MAD models at varying viewing angles (0$^\circ$ vs. 90$^\circ$) and frequencies (86 GHz vs. 230 GHz). The anisotropic size constraints are $5.44M<\theta_\mathrm{min}<10.68M$ and  $9.87M<\theta_\mathrm{maj}<12.69M$ at 230 GHz and $16.5<\theta_\mathrm{min}<23.8M$ and $17.3M<\theta_\mathrm{maj}<31.0M$ at 86 GHz \citep{Issaoun2019}.
    }
	\label{tab:86GHzVs230GHzAnd90DegVs0Deg}
	\begin{tabular}{lccccccr} 
		\hline
		Model & 
        && $\theta_\mathrm{min}(M)$
        && 
        & $\theta_\mathrm{maj}(M)$ 
        &  \\
        &&&&&&\\
     & 
     & 86 GHz, 0$^\circ$ &  86 GHz, 90$^\circ$  &  230 GHz, 0$^\circ$ & 86 GHz, 0$^\circ$ & 86 GHz, 90$^\circ$  & 230 GHz, 0$^\circ$   \\
     		\hline
	 Electron Evolution
          & 
          & 
          37.6  
          &
           25.7  
          & 
          26.8  
          & 
          39.7
          & 
          47.8  
          & 
          28.5  
          \\
		\hline
		Electron Temperature ($f,\beta_c$) &    & & & \\
        ($0.1,0.01$)& 
        & 
        43.7  
        &   
         30.4   
        & 
        33.8  
        & 
        54.7  
        & 
        69.7  
        & 
        39.1     
        \\
       ($0.1,0.1$)&  
       & 
       39.8  
       &  
       29.6   
       & 
       28.9 
       & 
       48.7  
       &  
       62.5 
       & 
       32.2  
       \\
       ($0.1,1.0$)&  
       & 
       50.6  
       &  
       29.6  
       & 
       25.9   
       & 
       45.5  
       &   
       56.4 
       &  
       28.6  
       \\ 
       ($0.5,0.01$) & 
       & 
       42.5   
       &   
       28.4   
       &  
       34.9    
       &  
       47.6  
       &   
       59.5   
       &  
       39.5  
       \\
       ($0.5,0.1$)&  
       & 
       36.6  
       &  
       28.7  
       & 
       25.7 
       &  
       38.4 
       &  
       38.6 
       & 
       27.2 
       \\
       ($0.5,1.0$)&  
       & 
       33.9  
       & 
       27.6   
       & 
       24.3 
       &  
        34.8 
       & 
       \textbf{29.4} 
       & 
       25.4 
       \\ 
        \hline
		Constant Electron Beta ($\beta_{e0}$)&  &  & & &\\
         0.01   &  
         & 
         44.1  
         &  
         32.9  
         & 
         40.1   
         & 
        48.2
         & 
         49.9    
         & 
         42.1  
         \\
         0.1   & 
         & 
         56.6  
         &  
         45.8  
         & 
         54.0 
         & 
         59.5 
         & 
         50.3  
         & 
         57.6 
         \\ 
         1.0   & 
         &
         61.5  
         &  
         45.7  
         & 
         61.5   
         & 
        62.8 
         & 
         57.0 
         & 
         62.9  
         \\
        \hline
		 Magnetic Bias ($n$)
         &  &  & & & \\
		0 &  
        & 
        64.5  
        &  
        58.8 
        & 
        63.5  
        & 
        66.9 
        & 
        62.0  
        & 
        66.2 
        \\
        1 & & 
        61.5  
         &  
         45.7  
         & 
         61.5  
         & 
           62.8 
         & 
         57.0 
         & 
         62.9    
        \\
        2 & 
        & 
        60.7  
        & 
        44.4 
        & 
        58.9 
        & 
        62.6  
        & 
        55.0  
        & 
        61.1  
        \\ 
        \hline
	\end{tabular}
\end{table*}\label{AnisotropicGaussianSizeChangingThetaAndNu}

\subsection{Comparison of Spectra}

The 
Electron Evolution Model with Turbulent Heating fits most of the data near the sub-mm bump. Spectra in models inspired by equipartition (Constant $\beta_e$ and Bias) are dominated by near-horizon outflow emission (this becomes more apparent in images at lower frequencies, as the lensed disk becomes less prominent). These spectra significantly overproduce frequencies above infrared. 
For the Bias Model, lower $n$ reduces the falloff of $u_e$ with radius, accounting for flatter radio and X-ray spectra.  Spectra in the Critical Beta Electron Temperature Model tend to be dominated by outflow or coronal emission, especially at lower $f$ or $\beta_c$, 
and tend to be more peaked than the data. 


The following summarizes trends appearing upon comparing spectra in different models:
\begin{itemize}
\item{The Critical Beta Electron Temperature Model reproduces low- and high-energy spectral amplitudes over the $10^{11}$-$10^{19}$ Hz frequency domain better than the Constant $\beta_e$ and Bias Models.}
\item{The radio-IR spectrum is flatter for lower $\beta_{e0}$ in the Constant $\beta_e$ Model.}
\item{The radio-IR spectrum is flatter for lower $n$ in the Bias Model (until a sharp low frequency turnover at $\lesssim10^{11}$ Hz, where lower $n$ models steepen).}

\item{Spectra in the 
Constant $\beta_e$ and Bias Models appear flatter than spectra in the Critical Beta Electron Temperature Model.
}

\end{itemize}
Some of these trends 
are a consequence of varying 
mass accretion rate and synchrotron absorption
in different models, as discussed below.

\subsubsection{Trends With Mass Accretion Rate}

From Table \ref{MUnit230GHz}, 
model families with flatter spectra (Constant $\beta_e$ and Bias Models) tend to have lower (magnitude) mass accretion rates than the family with steeper spectra (the Critical Beta Electron Temperature Model). This can be explained from the simulation as follows: As $|\dot{m}|$ decreases at fixed flux at 230 GHz, 
temperature increases and plasma in regions that have not previously been emitting synchrotron radiation begin to emit, broadening the range of emitting temperatures and, in turn, the spectra. This also predicts an 
anticorrelation between the magnitude of the mass accretion rate and the size of the emitting region in images in optically thin synchrotron models, but optically thick models   
deviate markedly from this trend. 

\subsection{Phenomenological Classification}

For our fiducial semi-MAD simulation, upon exploration of 
parameter space, $(f,\beta_c)$ models tend to outperform the jet-inspired Constant $\beta_e$ and Bias $(n)$ Models with respect to 
spectral observations, and often outperform the detailed Electron Evolution Model with Turbulent Heating as well.



With respect to image morphology, observations favor compactness. 

The images best satisfying the  86 GHz GMVA and 230 GHz EHT elliptical Gaussian size constraints for $\theta_\mathrm{maj}$ are from the $(f,\beta_c)=(0.5,1)$ model. It is noteworthy that this highest $(f,\beta_c)$ pair producing the best spatial fit for the Critical Beta Electron Temperature Model also provides the best overall spectral fit, giving us indication that the image and spectral properties are correlated. Moreover, image compactness is related to image shape in our models, as the lowest $\beta_{e0}=0.01$ providing the best (smallest) emitting region size for equipartition-inspired models has image morphology resembling the asymmetric crescent from the $(f ,\beta_c )=(0.5,1)$ Model.



Upon a scan of model parameter space for our fiducial  simulation, we find that images tend to aggregate into (at least) 4 broad categories, which are closely tied to 
spectra  
and related to variability as well:
\renewcommand{\theenumi}{\Roman{enumii}}
\begin{enumerate}
    \item I.) Thin, Compact, Asymmetric Photon Ring/Crescent\newline
  This image morphology is exhibited for the Electron Evolution Model with Turbulent Heating and in parameterized models with high $(f,\beta_c)$, e.g., $\{(0.5,0.1),(0.5,1)\}$ or low $\beta_{e0}$, e.g.,  0.01. The concomitant spectra are the best fit across our models, with the largest deviation coming from the $\beta_{e0}=0.01$ Model's very flat spectrum in the IR band and mild X-ray excess. The mm-variability is characterized by light curves with moderate or high amplitude, oscillating rapidly on intra-hour scales-- with the greater variability in the Constant $\beta_{e0}$ branch breaking image degeneracy with $(f,\beta_c)$ models.
  \item II.) Inflow-Outflow Boundary + Thin Photon Ring\newline
  This image morphology is exhibited in models with $(f,\beta_c)\in \{(0.1,0.01),(0.1,0.1),(0.1,1),(0.5,0.01)\}$, and is accompanied by the steepest spectra, sharply peaking near the IR bump. The radio variability is moderate amplitude and fast. 
  \item III.) Thick Photon Torus
  \newline
  This image morphology is exhibited in the Constant $\beta_e$ Model with $\beta_{e0}\in \{0.1,1\}$, and the Bias Model with $
  n=2$, and is accompanied by spectra with large X-ray excesses (and a flat overall spectrum for $\beta_{e0}=0.1$). These models have low amplitude, slowly oscillating radio light curves. 
  \item IV.) Extended Outflow\newline
  Occurring in the $
  n=0$ Bias Model, this image morphology is linked to a flat overall spectrum with X-ray excess. This class is characterized by low amplitude, nearly monotonic radio variability within a day.
  \end{enumerate}

It is noteworthy that dominant image features are closely linked with dominant spectral features and tied to variability as well, and that these associations form the basis for distinct classes into which models governed by related physics-- albeit different parameterizations thereof-- can be identified. 

\subsubsection{SANE vs. SEMI-MAD vs. MAD Simulation Comparison}
 
As we consider simulations outside of the fiducial simulation, new image morphologies emerge. For example, for the SANE simulation, the $(f,\beta_c)=(0.1,1)$ Model has a distinctly helical outflow intensity map and a remarkably good spectral fit for all but the lowest frequencies (cf. 
Figs. \ref{fEqPt1orPt5ElectronTemperatureModelSANEsemiandMAD} and \ref{SpectrafEqPt1orPt5ElectronTemperatureModelSANEsemiandMAD}). 
Furthermore, it is notable that some of our models become similar to our favored $(f,\beta_c)=(0.5,1)$ Model in appropriate limits. In particular, as $\beta_{e0}$ goes from 1 to 0.01 in the Constant $\beta_e$ Model
, intensity maps become more confined to small cylindrical radius and asymmetric for SANE, semi-MAD and MAD cases alike; and, for the MAD simulation, closer spectral fits are produced for the microwave, IR and (especially) X-ray bands. 
The theoretical advancement of a single model unifying emission characteristics of accretion flow, corona and outflow using a single simulation and small set of parameters would enable us to directly identify observed features with plasma and emission physics of different AGN components. 

 


\section{Conclusions and Future Directions}


 We have used two simple classes of parametric emission prescriptions-- the 
turbulent-heating-based Critical Beta  Electron Temperature 
Model $T_e=fT_pe^{-\beta/\beta_c}$ and the equipartition-based Constant Electron Beta/Magnetic Bias Models 
$P_e=K_nP_B^n,\ (\beta_{e0}\equiv K_1\ \mathrm{for}\ n=1)$-- to 
explore a wide range of possible models for the images and spectra 
in the inner tens of gravitational radii around the supermassive black hole at the Galactic Center. One of our models, $(f,\beta_c)=(0.5,1)$, 
is observationally favored for Sgr A* due to its agreement with respect to the spectrum, emitting region compactness and asymmetry. We stress that both observationally preferred images and spectra are associated with emission concentrated in a bright, crescent-shaped portion of the photon ring near the horizon, and that forthcoming EHT and GRAVITY data will further constrain our models. 

It is worth noting that our intuitive one- and two-parameter models span new 
electron physics beyond what has been previously explored, and promise to bear upon AGN beyond Sgr A*. By surveying synthetic images and spectra in other parts of model parameter space, we may isolate the emission physics underlying 
particular observational phenomena. We summarize these results as follows. 

\subsection{Images Summary}

Synthetic 230 GHz intensity maps on the scale of tens of gravitational radii appear: 

\begin{itemize}
\item{Dominated by the inflow/outflow interface for the Critical Beta Electron Temperature Model }
\item{More compact and asymmetric with increasing $f$ or $\beta_c$ }
\item{Mixed with outflow/near-horizon photon ring emission for 
the Constant $\beta_e$ and Bias Models}
\item{More concentrated around  
horizon-circulating lensed disk emission} for increasing $\beta_{e0}$ and increasing $n$
\end{itemize}

\subsection{Spectra Summary}

In comparison to the data, our parameterized synthetic $10^{11}$ Hz $<\nu<10^{19}$ Hz spectra have:
\begin{itemize}
\item{More peaked slopes fairly consistent with the sub-mm bump but underproducing the low frequency tail for the Critical Beta Electron Temperature Model}
\item{Fairly consistent low frequency slope but flatter peak and overproduction of low and (especially) high frequency emission for the Constant $\beta_e$ Model}
\item{Consistent low frequency slope but broader peak and (in some cases vast) overproduction of high frequency emission for the Bias Model}

\end{itemize}


The steep radio spectra seen in the Electron Temperature Model is characteristic of emission from an adiabatically expanding coronal outflow, in which temperature rapidly declines in radius. The flatter radio spectra from the Constant $\beta_e$ and Bias prescriptions are due to contributions from the highly magnetized outflow and the dependence of the latter emission functions exclusively on $P_B$. Although these latter models can explain many features of jets/outflows (including Sgr A*) such as Doppler beaming, knots and other magnetic field substructure, we have shown that they do not accurately describe the inner regions of discs/jets around Sgr A* (and perhaps, by analogy, other systems).

\subsection{Future Directions}

Sgr A* viewed at 1.3 mm has exhibited intrahour variability \citep{Johnson2015} in the inner  $\sim 6r_g$ around a black hole whose Schwarszchild-radius-light-crossing-time is 40 s. 
In the future, it would be valuable to produce models with light curves in the simulations closely replicating the observational cadence and understand which range of models is most consistent with the observed variability. We may also add polarization maps to our pipeline to test whether electromagnetically dominated emission models in simulations with ordered magnetic field substructure, e.g., helical Blandford-K\"onigl outflows, can 
help explain observations of a high degree of linear polarization in AGN cores. 
     
The \say{observing} simulations methodology is a key link between ever-advancing simulations and observations of the central engines of JAB systems. The EHT serves as a timely testbed for our 
emission models that aim to unite simulations and observations. The next works planned in this ```Observing' JAB Simulations" series are applications to the prominent jets in the giant elliptical galaxy M87 and the highly variable quasar 3C 279. Other EHT target sources to be observed-- and possibly, \say{observed}-- in the future include: Cen A, NGC 1052 and OJ 287.

\section*{Acknowledgements}


This work was supported in part by NSF grants AST 13-33612 and AST 1715054, Chandra theory grant TM7-18006X from the Smithsonian Institution, and a Simons Investigator award from the Simons Foundation. This work was made possible by computing time granted by UC Berkeley on the Savio cluster. RA carried out part of this work supported by the California Alliance and Simons Foundation and the remainder of this work at the Black Hole Initiative at Harvard University, which is supported by a grant from the John Templeton Foundation. RA is also supported by the NSF MSIP Grant AST-1440254. SR was supported in part by the NASA Earth and Space Science Fellowship and the Gordon and Betty Moore Foundation
through Grant GBMF7392 during part of the duration of this work. We all thank the scientific editor's thorough and incisive comments, which greatly improved the overall quality of this work.




\bibliographystyle{mnras}
\bibliography{ARQ20} 




\appendix
\section[TablesAppendix]{Tables}\label{TablesAppendix}

\subsection{Magnetic Bias Model Normalization Constant}

For the Bias Model, the electron gas pressure $P_e$ is prescribed to scale as powers $n$ of the magnetic pressure $P_B$, requiring a magnetic-to-gas pressure conversion factor with units of pressure to power $1-n$. Table \ref{OldAvgbToNCylinder} provides this factor by imposing a normalization condition in the semi-MAD simulation that the numerical average of $P_B^2$ equals that of $K_nP_B$ in the region between concentric cylindrical surfaces at $R=r_\rho=1.89M
$ and $R=20M$, as well as $50M$ for 
comparison. An alternative normalization region 
 in which the averaged is take over annuli on the equatorial plane with the height of one simulation pixel produces similar results. Since the averages are similar for our different choices of region geometry and extent, we consider our normalization approach robust.

\subsection[StatisticalAnalysis:Moments]{Statistical Analysis: Moments}\label{StatisticalAnalysis:Moments}


We may make quantitative our comparison of the disparate array of images generated from distinct physical processes by a comparison of statistical moments: 
\begin{equation}
M_{n_1,n_2}=\sum_{i,j}x_i^{n_1}y_j^{n_2}I(x_i,y_j)
\end{equation}
First and second order image moments (centroid and gyroradius) for $-50M<x,y<50M$ 230 GHz images in key models are presented in Table \ref{tab:ImageMoments230GHz}. The centroid of image intensity is right-offset, arising from asymmetric Doppler boosting of the accretion flow. However, it remains within the narrow band between $5M<x<10M$ and $-M<y<3M$. The centroid right offset is consistent with the apparent mirror asymmetry in synthetic image brightness, especially in models producing a prominent bright spot on the inner right edge of the disk. The higher moment gyroradius, including standard deviation, has greater variation across models due to contributions from larger radii emitting segments generated from low $n$ or $\beta_{e0}$ portions 
of Constant-$\beta_e$/Bias Model parameter space.

The image covariance matrices built from these moments are presented in Tables \ref{AnisotropicGaussianSize} and \ref{AnisotropicGaussianSizeChangingThetaAndNu} in the main text. 

\newpage
\begin{table}
	\centering
	\caption{Average values $\langle b^N\rangle$ from cylindrical radii $R=r_\rho=1.87M
	$ to $R_\mathrm{max}\in\{20M,50M\}$.} 
	\label{OldAvgbToNCylinder}
    \begin{tabular}{lccr} 
		\hline
		$N$ & Avg. out to $R_\mathrm{max}=20M$ & Avg. out to $R_\mathrm{max}=50M$ \\
     && \\
		\hline 
		1 & 0.0064
		& 0.0023
		\\
		2 & 0.00011
		& 1.75e-05 
		\\ 
		3 & 5.52e-06
		& 7.22e-07 
		\\
        4 & 4.63e-07
        & 5.95e-08 
        \\ 
        5 & 4.57e-08
        & 5.87e-09
        \\
        6 & 4.89e-09
        & 6.29e-10
        \\
		\hline
	\end{tabular}
\end{table}
 

\begin{table*}
	\centering
	\caption{Images moment comparison of $1^\mathrm{st}$ and $2^\mathrm{nd}$ moments for $-50M<x,y<50M$ at 230 GHz for key models.
}
	\label{tab:ImageMoments230GHz}
	\begin{tabular}{lccr} 
		\hline 
		Model & Centroid $\left(\frac{M_{10}}{M_{00}},\frac{M_{01}}{M_{00}}\right)$ & Radius of Gyration $\left(\sqrt{\frac{M_{20}}{M_{00}}},\sqrt{\frac{M_{02}}{M_{00}}},\sqrt{\frac{M_{20}+M_{02}}{M_{00}}}\right)$   \\
     		\hline
		Electron Evolution with Turbulent Heating  
          & $(6.56M,-0.15M)$
          & $(9.96M, 11.56M, 15.26M)$
          \\
		\hline
		Critical Beta Electron Temperature ($f,\beta_c$) &  &  \\
        ($0.1,0.01$) & $(5.18M, 1.41M)$
        & $(11.76M, 20.58M, 23.70M)$
        \\
       ($0.1,0.1$) & 
       $(5.25M, 1.65M)$
       & $(10.70M, 16.92M, 20.02M)$
       \\
       ($0.1,1.0$)& $(5.45M, 1.40M)$
       & $(10.39M, 14.31M, 17.69M)$
       \\
($0.5,0.01$) & $(6.36M, 2.91M)$
& $(11.51M, 18.80M, 22.04M)$
\\
       ($0.5,0.1$)& 
       $(8.13M, 1.80M)$
       & $(10.81M, 10.53M, 15.09M)$
       \\
       ($0.5,1.0$)& $(8.61M, 1.17M)$
       & $(10.78M, 7.93M, 13.38M)$
       \\
        \hline
		Constant Electron Beta ($\beta_{e0}$)&  &   \\
         0.01   & $(9.02M, 1.92M)$
         & $(16.44M, 21.41M, 26.99M)$
         \\
         0.1   & $(8.47M, -0.65M)$
         & $(24.51M, 22.55M, 33.31M)$
         \\
         1.0   & $(5.32M, 0.24M)$
         & $(27.32M, 23.24M, 35.86M)$ 
         \\ 
        \hline
		 Magnetic Bias ($n$)
         &  &   \\
		0 & $(5.50M, -0.14M)$
        & $(26.44M, 30.05M, 40.02M)$
        \\
        1 & $(5.32M, 0.24M)$
        & $(27.31M, 23.24M, 35.86M)$
        \\
        2 & $(5.76M, -0.62M)$
        & $(26.85M, 22.86M, 35.26M)$ 
        \\
        \hline
	\end{tabular}
\end{table*}

 \begin{table*} 
	\centering
	\caption{Pass-fail table of fiducial (edge-on, 230 GHz, semi-MAD simulation) models with respect to spectrum and all models for morphology (i.e., compactness/asymmetry measured by anisotropic Gaussian characteristic size $\theta_\mathrm{maj}$). Passing spectra intersect at least one data point in the frequency band considered.  Failing spectra overproduce emission.  F* indicates models that underproduce emission in the IR or X-Ray bands where we expect non-thermal contributions neglected in our calculation to be important). For morphology, the simulations (SANE (S), semi-MAD (sM) and/or MAD (M)), observing angles and frequencies are specified for passing models.} 
	\label{PassFail}
	\begin{tabular}{lccccr} 
		\hline
		Model & 
       Radio &  NIR & X-Ray & Morphology & Morphology
        \\
     & Spectrum 
     & Spectrum
     & Spectrum & 86 GHz & 230 GHz
     \\
     		\hline
		Electron Evolution with Turbulent Heating  
          & \textbf{P}
          & \textbf{P}
          & F
          & F
          & F
          \\
		\hline
		 Critical Beta Electron Temperature ($f,\beta_c$) &  
        & \\
        ($0.1,0.01$)           & \textbf{P}
          & F*
          & F*
          & \textbf{P} (S/$0^\circ$; M/$0^\circ$)
          & F
        \\
       ($0.1,0.1$)& \textbf{P}
          & F*
          & F*
          & \textbf{P} (S/$0^\circ$; M/$0^\circ$)
          & F 
       \\
       ($0.1,1.0$)& 
       \textbf{P}
          & F*
          & F*
          & \textbf{P} (S/$0^\circ$; M/$0^\circ$) 
          & F 
       \\
       ($0.5,0.01$)& \textbf{P}
          & F*
          & F*
          & \textbf{P} (S/$0^\circ$; M/$0^\circ$) 
          & F
       \\
       ($0.5,0.1$)& \textbf{P}
          & \textbf{P}
          & F*
          & \textbf{P} (S/$0^\circ$; M/$0^\circ$) 
          & F
       \\
       ($0.5,1.0$) & \textbf{P}
          & \textbf{P}
          & \textbf{P}
          & \textbf{P} (S/$0^\circ$; S/$90^\circ$; M/$0^\circ$;   
          & \textbf{P} (S/$90^\circ$)
       \\
       &&&& 
        M/$90^\circ$; 
        sM/$90^\circ$)& \\
        \hline
		Constant Electron Beta ($\beta_{e0}$)&  &  
		\\
         0.01   & \textbf{P}
          & \textbf{P}
          & F
          & \textbf{P}  (M/$0^\circ$)
          & F 
         \\
         0.1   & \textbf{P}
          & \textbf{P}
          & F
          & \textbf{P}  (M/$0^\circ$)
          & F
         \\
         1.0   & \textbf{P}
          & F
          & F
          & \textbf{P}  (M/$0^\circ$)
          & F
         \\
        \hline
		 Magnetic Bias ($n$)
         &  &  
         \\
		0 & F
          & F 
          & F
          & \textbf{P}  (M/$0^\circ$)
          & F
        \\
        1 & \textbf{P}
          & F
          & F
          & \textbf{P}  (M/$0^\circ$)
          & F
        \\
        2 & \textbf{P}
          & F
          & F
          & \textbf{P}  (M/$0^\circ$)
          & F
        \\
        \hline
	\end{tabular}
\end{table*}\label{PassFailTable}

\bsp	
\label{lastpage}
\end{document}